\documentclass[aps,onecolumn,floatfix,altaffilletter,preprintnumbers,
               tightenlines,showpacs,showkeys,notitlepage,nofootinbib]{revtex4-2}
\usepackage{amsmath}
\usepackage{amssymb}
\usepackage{titlesec}
\usepackage[a4paper,width=165mm,top=25mm,bottom=25mm]{geometry}
\usepackage{slashed}
\usepackage{braket}
\usepackage{bbm}
\usepackage[capitalise]{cleveref}
\usepackage{xcolor}
\usepackage{graphicx}
\usepackage[caption=false,labelformat=simple]{subfig}
\usepackage{siunitx}
\usepackage[force]{feynmp-auto}

\begin{document}

\title{A Scotogenic Model as a Prototype for Leptogenesis with One Single Gauge Singlet}

\author{Björn Garbrecht}
\email{garbrecht@tum.de}
\affiliation{Physik-Department T70, Technische Universität München, James-Franck-Straße, 85748 Garching, Germany}

\author{Edward Wang}
\email{edward.wang@tum.de}
\affiliation{Physik-Department T70, Technische Universität München, James-Franck-Straße, 85748 Garching, Germany}

\preprint{TUM-HEP-1505/24}

\begin{abstract}
\noindent
We investigate the potential of a minimal scotogenic model with two additional scalar doublets and a single heavy Majorana fermion to explain neutrino masses, dark matter, and the baryon asymmetry of the Universe. In this minimal setup, leptogenesis is purely flavored, and a second Majorana neutrino is not necessary because the Yukawa couplings of the extra doublets yield the necessary $CP$-odd phases. The mechanism we employ can also be applied to a wide range of scenarios with at least one singlet and two gauge multiplets. Despite stringent limits from the dark matter abundance, direct detection experiments, and the baryon asymmetry of the Universe, we find a parametric region consistent with all bounds which could resolve the above shortcomings of the Standard Model of particle physics. Methodically, we improve on the calculation of correlations between the mixing scalar fields given their finite width. We also present an argument to justify the kinetic equilibrium approximation for out-of-equilibrium distribution functions often used in calculations of baryogenesis and leptogenesis.
\end{abstract}

\maketitle

\section{Introduction}

Leptogenesis is a possible explanation of the baryon asymmetry of the universe (BAU) motivated by neutrino mass mechanisms. In the type-I seesaw mechanism, left-handed neutrinos couple to at least two heavy Majorana fermions through the Standard Model (SM) Higgs doublet, so that at least two neutrinos become massive. The interferences between amplitudes involving the different Majorana fermions give rise to $CP$-odd phases in their decays \cite{Fukugita:1986hr}. The asymmetry arising from these $CP$-violating decays is first produced in the leptons and subsequently transferred to the baryon sector via sphaleron transitions. In the present work we explore an alternative mechanism, in which $CP$-violation arises from interferences of amplitudes involving different multiplets of the SM gauge group instead of singlets, building upon previous work in Refs.~\cite{Garbrecht:2012qv,Garbrecht:2012pq}. A singlet however still induces the necessary deviation from equilibrium, but for this sole purpose, just a single one is sufficient. When the multiplets taking part in the interferences do not have lepton number violating interactions, leptogenesis is purely flavored, i.e., the net sum of the decay and inverse decay asymmetries over the flavors is zero. Nonvanishing symmetries in the particular flavors lead to a net unflavored asymmetry through washout processes involving the singlet. Minimal scenarios therefore consist of one singlet and at least two multiplets leading to mixing and interference, and the latter must couple to at least two different flavors of SM fermions.

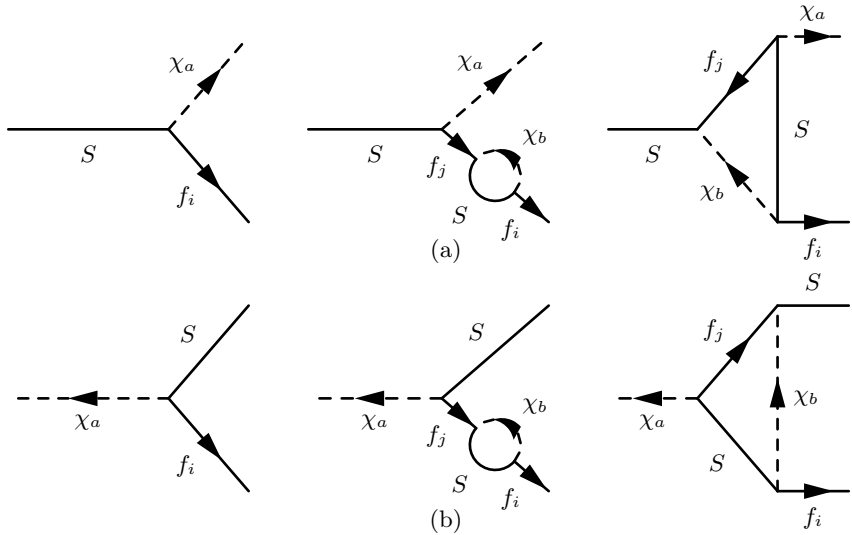
\begin{figure}
\subfloat[]{
\begin{fmffile}{tree-level_1}
\begin{fmfgraph*}(100,70)
\fmfleft{i}
\fmfright{o1,o2}
\fmf{plain, label=$S$}{i,v1}
\fmf{scalar, label=$\chi_a$, label.side=left}{v1,o2}
\fmf{fermion, label=$f_i$, label.side=right}{v1,o1}
\end{fmfgraph*}
\end{fmffile}
\begin{fmffile}{CP-source_wf_1}
\begin{fmfgraph*}(100,70)
\fmfleft{i}
\fmfright{o1,o2}
\fmf{plain, label=$S$}{i,v1}
\fmf{scalar, label=$\chi_a$, label.side=left}{v1,o2}
\fmf{phantom}{v1,o1}
\fmffreeze
\fmfshift{(-10,0)}{v1}
\fmf{phantom}{v1,v2,v3,o1}
\fmf{fermion, label=$f_j$}{v1,v2}
\fmf{plain, right, tension=0.4, label=$S$}{v2,v3}
\fmf{scalar, right, tension=0.4, label=$\chi_b$}{v3,v2}
\fmf{fermion, label=$f_i$, label.side=right}{v3,o1}
\end{fmfgraph*}
\end{fmffile}
\begin{fmffile}{CP-source_v_1}
\begin{fmfgraph*}(100,70)
\fmfleft{i1,i,i2}
\fmfright{o1,o3,o2}
\fmffreeze
\fmf{phantom}{i2,v5,v2,o2}
\fmf{phantom}{i1,v6,v3,o1}
\fmf{phantom}{i,v1,v4,o3}
\fmffreeze
\fmf{plain, label=$S$}{i,v1}
\fmf{scalar, label=$\chi_b$}{v3,v1}
\fmf{fermion, label=$f_j$, label.side=right}{v2,v1}
\fmf{plain, label=$S$}{v3,v2}
\fmf{scalar, label=$\chi_a$, label.side=left}{v2,o2}
\fmf{fermion, label=$f_i$}{v3,o1}
\end{fmfgraph*}
\end{fmffile}
\label{subfig:singlet_decay}} \par \bigskip
\subfloat[]{
\begin{fmffile}{tree-level_2}
\begin{fmfgraph*}(100,70)
\fmfleft{i}
\fmfright{o1,o2}
\fmf{scalar, label=$\chi_a$, label.side=left}{v1,i}
\fmf{plain, label=$S$, label.side=left}{v1,o2}
\fmf{fermion, label=$f_i$, label.side=right}{v1,o1}
\end{fmfgraph*}
\end{fmffile}
\begin{fmffile}{CP-source_wf_2}
\begin{fmfgraph*}(100,70)
\fmfleft{i}
\fmfright{o1,o2}
\fmf{scalar, label=$\chi_a$, label.side=left}{v1,i}
\fmf{plain, label=$S$, label.side=left}{v1,o2}
\fmf{phantom}{v1,o1}
\fmffreeze
\fmfshift{(-10,0)}{v1}
\fmf{phantom}{v1,v2,v3,o1}
\fmf{fermion, label=$f_j$}{v1,v2}
\fmf{plain, right, tension=0.4, label=$S$}{v2,v3}
\fmf{scalar, label=$\chi_b$, right, tension=0.4}{v3,v2}
\fmf{fermion, label=$f_i$, label.side=right}{v3,o1}
\end{fmfgraph*}
\end{fmffile}
\begin{fmffile}{CP-source_v_2}
\begin{fmfgraph*}(100,70)
\fmfleft{i1,i,i2}
\fmfright{o1,o3,o2}
\fmffreeze
\fmf{phantom}{i2,v5,v2,o2}
\fmf{phantom}{i1,v6,v3,o1}
\fmf{phantom}{i,v1,v4,o3}
\fmffreeze
\fmf{scalar, label=$\chi_a$, label.side=left}{v1,i}
\fmf{plain, label=$S$}{v3,v1}
\fmf{fermion, label=$f_j$, label.side=left}{v1,v2}
\fmf{scalar, label=$\chi_b$}{v3,v2}
\fmf{plain, label=$S$, label.side=left}{v2,o2}
\fmf{fermion, label=$f_i$}{v3,o1}
\end{fmfgraph*}
\end{fmffile}
\label{subfig:multiplet_decay}}
\caption{Tree-level, wavefunction and vertex-type contributions to decays in the case of (a) singlet and (b) multiplet decay in a general class of models, where $f_{i,j}$ are SM fermions, $S$ is a singlet and $\chi_{a,b}$ are multiplets of the SM gauge group. While the dashed and solid lines typically represent scalars and fermions respectively, exchanging the fermionic/bosonic natures of $\chi$ and $S$ is also possible.}
\label{fig:decays}
\end{figure}

The interferences between amplitudes involving the different multiplets can then lead to an asymmetry in the decay of the heaviest particle, which can be either the singlet or one of the multiplets. In both cases, it is the singlet that drives the system out of equilibrium. Since the multiplets participate in the SM gauge interactions, they tend to equilibrate quickly and through processes that do not involve asymmetry production, whereas the singlet can only equilibrate via its interactions with the multiplets, where $CP$-violation occurs. The diagrams for the relevant tree and loop amplitudes are shown in \Cref{fig:decays}, where $S$ stands for the singlet and $\chi_{a,b}$ are the multiplets.

In order to preserve gauge invariance, the multiplets must have the same quantum numbers as the SM fermions to which they couple with the singlet at the renormalizable level. Examples of models containing additional $SU(2)_L$ doublets that can generally lead to $CP$-violating interferences are scotogenic models \cite{Hagedorn:2018spx,Kumar:2019tat,Escribano:2020iqq,Leite:2020bnb,Cacciapaglia:2020psm,Sarazin:2021nwo,Ahriche:2022bpx,Ghosh:2022fws,Alvarez:2023dzz,Bonilla:2023aij,Escribano:2023hxj,Karan:2023adm}, two-Higgs-doublet models \cite{Maniatis:2006fs,Keus:2013hya,Li:2014fea,Maniatis:2014oza,Chen:2014xva,Abbas:2015cua,Maniatis:2015kma,Han:2015yys,Hu:2016gpe,Hu:2017qxj,Abbas:2018pfp,Cogollo:2019mbd,Jurciukonis:2019jkr,Chen:2021jok,Wang:2021zfp,Enomoto:2022rrl,Sartore:2022sxh,Cai:2022xha,Connell:2023jqq,Karan:2023kyj,Darvishi:2023fjh,Arcadi:2023rbv}, and inert doublet models \cite{Machado:2012gxi,Fortes:2014dca,Keus:2014jha,Keus:2014isa,Alanne:2016wtx,Aranda:2019vda,Merchand:2019bod,Rosenlyst:2021tdr,Rosenlyst:2022jxj,Belanger:2022qxt,Khojali:2022squ,Singirala:2023zos}, and aim at addressing such problems as dark matter, neutrino masses, muon $g-2$, among others. Models containing additional scalar color triplets have also been proposed \cite{Angel:2013hla,Addazi:2015ata,Addazi:2015rwa,Carquin:2019xiz} and are a common prediction of grand unified theories, although there the mass of the triplet has to be very large to avoid proton decay, leading to the doublet-triplet mass splitting problem \cite{Dvali:1992hc,Dvali:1995hp,Kawamura:2000ev,vonGersdorff:2020lll}. Another possibility would be to have interferences involving sfermions or Higgsinos in supersymmetric models, which by construction have the same quantum numbers as their superpartners \cite{Pilaftsis:1999bq,Choi:2001ks,Li:2008kz,Kim:2008yta,Kim:2009nq,Cheung:2009qk,Baer:2011ec,Sakurai:2011pt,Kozaczuk:2012xv,Altmannshofer:2013lfa,Nagata:2013sba,Kim:2014noa,Maekawa:2017xci,Krall:2017xij,Arcadi:2022hve,Shafi:2023sqa}. The mechanism we lay out in the present work can by and large be applied to these scenarios, when one or more multiplets mix and interfere to produce $CP$-violating interactions. While in the remainder of this work we will consider the usual case of a fermionic singlet and a scalar multiplet, the opposite case is also possible.

In the present work, we demonstrate how this general mechanism applies to a minimal variant of the scotogenic model. scotogenic models are a class of beyond the Standard Model (BSM) scenarios aiming to explain the smallness of neutrino masses while also including a dark matter (DM) candidate. The original scotogenic model \cite{Ma:2006km} extends the SM by a dark sector, odd under a new $\mathbb{Z}_2$ symmetry and containing one extra Higgs doublet and two Majorana fermions. Due to the $\mathbb{Z}_2$ symmetry, the lightest particle of this new sector is absolutely stable and therefore a dark matter candidate. Soon after its proposal, it was realized that the scotogenic model also has the potential for producing leptogenesis in a similar way as in the Seesaw model \cite{Ma:2006fn}.

Several variants of the original model have been put forward and studied extensively. The model we investigate here was proposed in Ref.~\cite{Hehn:2012kz} and is an alternative minimal realization of the scotogenic model, with two additional scalar doublets instead of one and only a single Majorana fermion. The goal of this work, in addition to the above considerations, is to explore the possibility of explaining neutrino masses, the baryon asymmetry of the universe, and dark matter in this minimal scenario. Similar attempts, based on other variants of the scotogenic model, have been reported in Refs.~\cite{Baumholzer:2018sfb, Chen:2020ark, Sarma:2020msa, Mahanta:2019gfe, Borah:2018rca,Borah:2021qmi,Suematsu:2024rtu}.

The outline of the article is as follows: In \cref{sec:model} we present the model and its properties, in \cref{sec:constraints} we discuss some of the main constraints of the model and in \cref{sec:leptogenesis} we present the details of our calculation of leptogenesis. In \cref{sec:dark_matter} we discuss the dark matter production in our model and in \cref{sec:parameters} we present the allowed parameter region.

\section{The Model}
\label{sec:model}

The model we consider was proposed in Ref.~\cite{Hehn:2012kz} and extends the Standard Model by one Majorana fermion $N$ and two complex scalars $\eta_{1,2}$, doublets under $SU(2)_L$ and with hypercharge $1/2$. Furthermore, a discrete $\mathbb{Z}_2$ symmetry is imposed under which the Standard Model particles are even, while the new particles are odd. The Lagrangian for this model is given by
\begin{equation}
\mathcal{L} = \mathcal{L}_\text{SM} + \mathcal{L}_N + \mathcal{L}_\eta + \mathcal{L}_\text{int}^\text{fermion} + \mathcal{L}_\text{int}^\text{scalar},
\end{equation}
where $\mathcal{L}_\text{SM}$ is the SM Lagrangian. The scotogenic sector is introduced through
\begin{align}
\mathcal{L}_N &= \frac{1}{2} \bar{N}^c i \slashed{\partial} N - \frac{1}{2} M_N \bar{N}^c N + \text{h.c.}, \\
\mathcal{L}_\eta &= (D_\mu \eta^a)^\dag (D^\mu \eta_a) - (m_\eta^2)_{ab} \eta^{a \dag} \eta^b - V(\eta_1, \eta_2),
\end{align}
where $D_\mu$ is the covariant derivative and $V(\eta_1, \eta_2)$ is the general potential of a two-Higgs doublet model. We can define the mass matrix $(m_\eta^2)_{ab}$ to be diagonal, with values $m_{\eta 1}^2$ and $m_{\eta 2}^2$. Interactions between the SM and scotogenic sector are given by
\begin{align}
\mathcal{L}_\text{int}^\text{fermion} &= - Y_i^{(a)} \bar{N} (L_i i \sigma_2 \eta_a) + \text{h.c.} = - Y_i^{(a)} \bar{N} (\nu_i \eta_a^0 - l_i^- \eta_a^+) + \text{h.c.},\\
\mathcal{L}_\text{int}^\text{scalar} &= - \frac{1}{2} \lambda_3^{(ab)} (\Phi^\dag \Phi) (\eta_a^\dag \eta_b) - \frac{1}{2} \lambda_4^{(ab)} (\Phi^\dag \eta_a) (\eta_b^\dag \Phi) - \frac{1}{2} \lambda_5^{(ab)} (\Phi^\dag \eta_a) (\Phi^\dag \eta_b) + \text{h.c.},
\end{align}
where h.c. stands for Hermitian conjugation. We assume the discrete symmetry $\mathbb{Z}_2$ to remain unbroken, meaning that the fields $\eta_1$ and $\eta_2$ do not acquire a vacuum expectation value, whereas for the SM Higgs field $\Phi$ we have $\braket{\Phi^0} = v/\sqrt{2}$.

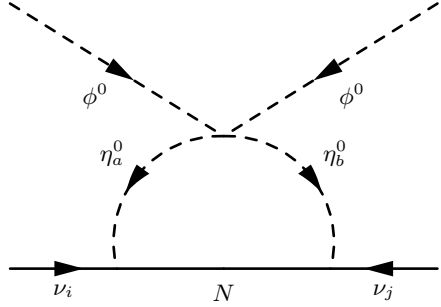
\begin{figure}
\begin{fmffile}{NeutrinoMass}
\begin{fmfgraph*}(200,100)
\fmfleft{i1,i2}
\fmfright{f1,f2}
\fmf{fermion, label=$\nu_i$, label.side=right}{i1,v1}
\fmf{fermion, label=$\nu_j$, label.side=left}{f1,v3}
\fmf{plain}{v1,v4,v3}
\fmffreeze
\fmfv{l=$N$,l.a=-90}{v4}
\fmf{scalar, right=0.5, label=$\eta_a^0$,label.side=right}{v2,v1}
\fmf{scalar, left=0.5, label=$\eta_b^0$}{v2,v3}
\fmf{scalar, label=$\phi^0$}{i2,v2}
\fmf{scalar,label=$\phi^0$}{f2,v2}
\end{fmfgraph*}
\end{fmffile}
\setlength{\abovecaptionskip}{16pt}
\caption{Diagram generating neutrino masses in this model.}
\label{fig:nu-mass}
\end{figure}

The left-handed neutrinos do not directly couple to the SM Higgs field and therefore cannot obtain mass at tree level. The leading contribution to the mass at one-loop order shown in \Cref{fig:nu-mass} is
\begin{equation}
(m_\nu)_{ij} = \frac{Y_i^{(a)} Y_j^{(b)} \lambda_5^{(ab)} v^2}{8 \pi^2} \frac{M_N}{m_{\eta_b}^2 - M_N^2} \left( \frac{m_{\eta_b}^2}{m_{\eta_a}^2 - m_{\eta_b}^2} \text{log} \left(\frac{m_{\eta_a}^2}{m_{\eta_b}^2} \right) - \frac{M_N^2}{m_{\eta_a}^2 - M_N^2} \text{log} \left( \frac{m_{\eta_a}^2}{M_N^2} \right) \right),
\label{eq:mass_matrix}
\end{equation}
where we sum over $a, b = 1,2$. Neutrino masses are then expected to be small if the couplings $Y$ and $\lambda_5$ are small or if the heaviest mass scale is much above the electroweak scale. In principle, different hierarchies of the masses of the new dark particles are possible; we will, however, restrict ourselves to the scenario in which 
\begin{equation}
	m_{\eta 2} > M_N > m_{\eta 1}.
\end{equation}

After electroweak symmetry breaking, the coupling of the new scalars to the SM Higgs field gives a correction to their masses. We can parametrize the scalar fields as
\begin{equation}
\eta_a = \begin{pmatrix}
\eta_a^+ \\
\eta_a^0
\end{pmatrix} =  \begin{pmatrix}
\eta_a^+ \\
\frac{1}{\sqrt{2}} (\eta_{R a} + i \eta_{I a})
\end{pmatrix},
\end{equation}
where $\eta_a^+$ is a charged scalar and $\eta_{R a}$ and $\eta_{I a}$ are $CP$ even and odd neutral real scalars, respectively. The new mass matrices accounting for the vacuum expectation value are
\begin{align}
(m_+^2)_{ab} &= (m_\eta^2)_{aa} \delta_{ab} + \lambda_3^{ab} \frac{v^2}{2}, \label{eq:masspl} \\
(m_R^2)_{ab} &= (m_\eta^2)_{aa} \delta_{ab} + \text{Re} (\lambda_3^{ab} + \lambda_4^{ab} + \lambda_5^{ab}) \frac{v^2}{2}, \label{eq:massR}\\
(m_I^2)_{ab} &= (m_\eta^2)_{aa} \delta_{ab} + \text{Re} (\lambda_3^{ab} + \lambda_4^{ab} - \lambda_5^{ab}) \frac{v^2}{2}.
\label{eq:massI}
\end{align}
In general, we assume these corrections $\sim \lambda v^2$ to be small compared to $m_\eta^2$. With this assumption, we can approximate the mass matrices as diagonal with eigenvalues $m_{Ra}, m_{Ia}$. The mass splitting between the neutral scalars of a doublet is then given by \cite{Escribano:2020iqq}
\begin{equation}
m_{Ra}^2 - m_{Ia}^2 = \text{Re} (\lambda_5^{aa}) v^2,
\label{eq:mass_splitting}
\end{equation}
with corrections appearing at second order in $\lambda v^2$.

\section{Constraints}
\label{sec:constraints}

\subsection{Lepton Flavor Violation}

Scotogenic models predict lepton flavor violating (LFV) processes such as the radiatively induced decays $\ell_i \to \ell_j \gamma$ and $\ell_\alpha \to 3 \ell_\beta$. The corresponding branching ratios have been computed in Ref. \cite{Toma:2013zsa}, and the relevant upper bounds from Ref. \cite{Workman:2022ynf} are listed in \cref{table:lfv}.

\begin{table}
\centering
\begin{tabular}{|c|c|}
\hline
Process & BR upper bound \\
\hline
$\mu^- \to e^- \gamma$ & $4.2 \times 10^{-13}$ \rule{0pt}{3.5mm} \\
$\tau^- \to e^- \gamma$ & $3.3 \times 10^{-8}$ \\
$\tau^- \to \mu^- \gamma$ & $4.2 \times 10^{-8}$ \\
$\mu^- \to e^- e^+ e^-$ & $1.0 \times 10^{-12}$ \\
$\tau^- \to e^- e^+ e^-$ & $2.7 \times 10^{-8}$ \\
$\tau^- \to \mu^- \mu^+ \mu^-$ & $2.7 \times 10^{-8}$ \\
\hline
\end{tabular}
\caption{LFV processes and their respective upper bounds, extracted from Ref. \cite{Workman:2022ynf}.}
\label{table:lfv}
\end{table}

\subsection{Direct detection}

The couplings of the dark matter particle $\eta_{1}$ to the SM Higgs and to the weak gauge bosons will produce a signature in direct detection experiments. If the $CP$ even and odd neutral components of the scalars are degenerate in mass, the spin-independent elastic cross section due to $Z$-boson exchange of the dark matter particle on nuclei is many orders of magnitude larger than allowed by experiments \cite{Barbieri:2006dq}. To avoid this constraint, it is necessary to have a sufficiently large mass splitting ($O(100)$ keV) between the two neutral scalars so that their kinetic energy is insufficient to upscatter in a ground-based detector. This can be achieved with sufficiently large $\lambda_5$, as per \cref{eq:mass_splitting}.

A second detection channel is through elastic scattering via Higgs boson exchange. This sets an upper bound on the allowed values for the scalar interactions. Defining $\lambda_{345} = \lambda_3 + \lambda_4 - |\lambda_5|$ as the coupling strength between the lightest dark scalar and the SM Higgs boson, the spin-independent DM-nucleon scattering cross section is given by
\begin{equation}
\sigma_\text{SI} = \frac{\lambda_{345}^2 f_n^2}{4 \pi} \frac{\mu^2 m_n^2}{m_h^4 m_\eta^2},
\end{equation}
with $\mu = m_n m_\eta/(m_n + m_\eta)$ the DM-nucleon reduced mass, $m_h$ the SM Higgs boson mass and $f_n \approx 0.32$ the Higgs-nucleon coupling \cite{Barbieri:1988zs}. This cross section is then constrained by direct detection experiments like LUX-ZEPLIN \cite{LZ:2022lsv}.

\subsection{Theoretical Constraints}

There are two main theoretically motivated constraints on our model. The first comes from the requirement of perturbativity. For a perturbative treatment of the theory to be possible, the coupling strengths should not be larger than $O(1)$. This is especially important since, as we will see, an interplay between the different masses and couplings is necessary to reproduce the correct neutrino masses, leading to unacceptably large couplings in some regions of the parameter space.

The second constraint comes from vacuum stability. The situation in three-Higgs-doublet models is similar to the two-Higgs-doublet case, which is well understood \cite{osti_5135337,Maniatis:2006fs,Branco:2011iw}, albeit somewhat more complicated. In general, however, the parameters of the scalar potential $V(\eta_1, \eta_2)$ can be chosen such that the full theory is stable. We will therefore not delve deeper into this issue, as this is not the focus of the present work. One important aspect, however, is that the dark matter candidate should be electrically neutral; this can be achieved if $\lambda_4 - |\lambda_5| < 0$, see \cref{eq:masspl,eq:massR,eq:massI}. For a more thorough discussion on vacuum stability in three-Higgs-doublet models see Refs. \cite{Keus:2013hya,Maniatis:2014oza,Maniatis:2015kma}.

\section{Leptogenesis}
\label{sec:leptogenesis}

In general, leptogenesis can be described by a set of coupled fluid equations for the particles under consideration, which are then used to track their evolution over time. To account for the expansion of the universe, it is convenient to write the kinetic equations in terms of yields $Y = g n/s$, where $g$ are the internal degrees of freedom of the field, $n$ is the particle number density for a single degree of freedom and $s$ is the entropy density. Since both particle and entropy densities are diluted with the expansion of the universe at the same rate (assuming no entropy is produced), this effect cancels out, and we do not need to include the Hubble rate in the kinetic equations explicitly. We further describe the evolution in terms of the following comoving dimensionful quantities: momentum $\vec{k} = a(t) \vec{k}_\text{phys}$, temperature $T = a(t) T_\text{phys}$ and entropy $s = a(t) s_\text{phys}$, where $a(t)$ is the scale factor from the Friedmann-Lemaître-Robertson-Walker metric. We label the corresponding physical quantities with the subscript phys. We work in conformal time $\eta$, which is related to the comoving time $t$ as $dt = a d \eta$, where, in a radiation-dominated universe, $a = a_R \eta$. For the comoving temperature and entropy density we set
\begin{equation}
a_R = T = \frac{m_{Pl}}{2} \sqrt{\frac{45}{g_\star \pi^3}}, \quad s = g_\star a_R^3 \frac{2 \pi^2}{45},
\end{equation}
where $m_{Pl} = 1.22 \times \SI{e19}{GeV}$ is the Planck mass and $g_\star = 114.75$ is the number of relativistic degrees of freedom with two additional Higgs doublets at high energies so that $\eta = 1/T_\text{phys}$. With this setup, the effect of the Hubble expansion on the scattering rates is captured by replacing all masses $m$ by $a(\eta) m$ in the rates that appear in the fluid equations.

For leptogenesis, this set of equations is given by
\begin{align}
\frac{d Y_N}{dz} &= \mathcal{C}_N (Y_N - Y_N^\text{eq}), \\
\frac{d Y_{\ell i}}{dz} &= S_{\ell i} (Y_N - Y_N^\text{eq}) + W_{\ell i} Y_{\ell i},
\label{eq:kinetic_ell}
\end{align}
with $Y_{\ell i} = g_w q_{\ell i}/s, Y_N = g_s n_N/s$, where $q_{\ell i}$ is the charge density of the leptons, $n_N$ is the number density of the Majorana fermion, and $g_s, g_w$ are the spin and $SU(2)$ degrees of freedom respectively. We use $z=M_N/T_\text{phys} = M_N \eta$ as a dimensionless time variable.

\begin{figure}
\captionsetup[subfloat]{captionskip=2em}
\subfloat[]{
\begin{fmffile}{CP-source_wf}
\begin{fmfgraph*}(200,70)
\fmfleft{i}
\fmfright{o}
\fmf{fermion, label=$\ell_i$}{i,v1}
\fmf{fermion, label=$\ell_i$}{v2,o}
\fmf{plain, right, tension=0.4, label=$N$}{v1,v2}
\fmf{dbl_dashes, left, tension=0.4, label=$\eta_{ab}$}{v1,v2}
\end{fmfgraph*}
\end{fmffile}
\setlength{\abovecaptionskip}{18pt}
\vspace{2em}
\label{subfig:wavefunction}}
\hspace{-1.5em}
\subfloat[]{
\begin{fmffile}{CP-source_vx}
\begin{fmfgraph*}(200,70)
\fmfleft{i1,i,i2}
\fmfright{f1,o,f2}
\fmf{phantom}{i1,v3,f1}
\fmf{phantom}{i2,v2,f2}
\fmffreeze
\fmf{fermion, label=$\ell_i$}{i,v1}
\fmf{fermion, label=$\ell_i$}{v4,o}
\fmf{plain, label=$N$,label.side=left}{v1,v2}
\fmf{scalar, label=$\eta_a$,label.side=left}{v3,v1}
\fmf{fermion, label=$\ell_j$}{v3,v2}
\fmf{plain, label=$N$}{v3,v4}
\fmf{scalar, label=$\eta_b$}{v4,v2}
\end{fmfgraph*}
\end{fmffile}
\setlength{\abovecaptionskip}{18pt}
\vspace{2em}
\label{subfig:vertex}}
\caption{Diagrammatic representation of the $CP$-violating wave function (a) and vertex-type (b) contributions to the source term. The double line for $\eta$ indicates the summation of the one-loop insertions, which allows for flavor correlations as indicated by the indices $ab$. Kinematic cuts of these diagrams produce diagrams of the form of \Cref{fig:decays}.}
\label{fig:CP-source}
\end{figure}
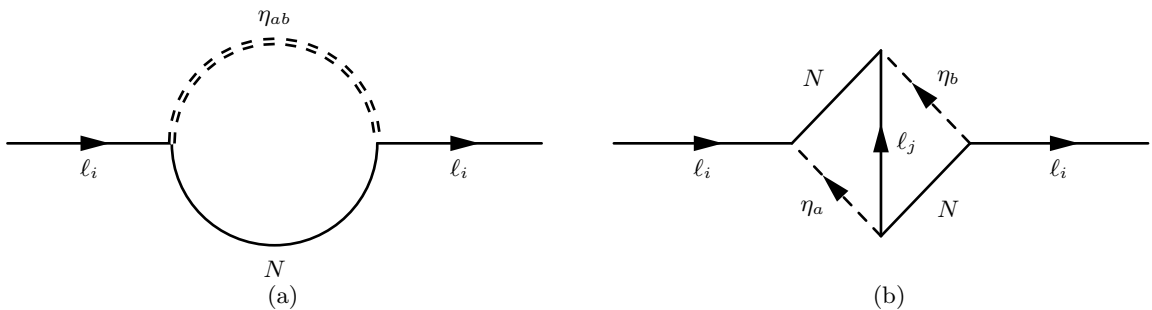

The $CP$-violating source term contains a wavefunction and a vertex-type contribution, $S_{\ell i} = S_{\ell i}^{\rm wf} + S_{\ell i}^{\rm v}$. In the closed-time-path (CTP) formalism, the wavefunction contribution to the $CP$-violating source is given by \cite{Garbrecht:2012qv}
\begin{equation}
S_{\ell i}^{\rm wf}´ = \sum_{a \neq b} Y_i^{(a) *} Y_i^{(b)} \int \frac{d^4 k}{(2 \pi)^4} \frac{d^4 p}{(2 \pi)^4} \frac{d^4 q}{(2 \pi)^4} (2 \pi)^4 \delta^4 (k + p - q) \text{tr} [i S_N^> (q) i S_{\ell i}^< (k) - < \leftrightarrow >] i D_{\eta ab} (p).
\end{equation}
The corresponding diagram is shown in \Cref{subfig:wavefunction}. The mixing scalar propagator that follows from the summation of all one-loop insertions is denoted by $D_\eta$. Its off-diagonal components can be obtained from the kinetic equation \cite{Garbrecht:2012qv,Garbrecht:2011aw, Garny:2011hg}
\begin{multline}
2 k^0 \partial_\eta i D_{\eta 12} + i (m_{\eta 1}^2 - m_{\eta 2}^2) i D_{\eta 12} = - \frac{1}{2} i(\Pi_{\eta 12}^{Y >} + \Pi_{\eta 12}^{\lambda >} + \Pi_{\eta 12}^{g >})i(\Delta_{\eta 11}^< + \Delta_{\eta 22}^<) \\
- \frac{1}{2} \sum_k i (\Pi_{\eta kk}^{Y >} + \Pi_{\eta kk}^{\lambda >} + \Pi_{\eta kk}^{g >}) i D_{\eta 12} - < \leftrightarrow >,
\label{eq:kinetic}
\end{multline}
where the $\Delta_\eta$ are the diagonal scalar propagators, which are assumed to be in thermal equilibrium, and $\Pi_\eta^Y, \Pi_\eta^\lambda, \Pi_\eta^g$ are the scalar self-energies arising from Yukawa, scalar and gauge interactions, respectively. The gauge and scalar interactions effectively bring the mixed propagator into kinetic equilibrium so that we can assume that the solutions to \cref{eq:kinetic} are of the form \cite{Garbrecht:2012qv,Beneke:2010dz}
\begin{equation}
i D_{\eta 12} (p) = 2 \pi \delta(p^2 - m_{\eta 1}^2) \frac{\mu_{\eta 12}}{T} \frac{\text{sign} (p^0) e^{|p^0|/T}}{(e^{|p^0|/T} - 1)^2},
\label{eq:propagator}
\end{equation}
with the chemical potential $\mu_{\eta 12}$. In principle, the mixed propagator should contain a second contribution with a pole in $m_{\eta 2}$ \cite{Garbrecht:2011aw}, however, since we assume $m_{\eta 2} \gg m_{\eta 1}$ we can neglect this contribution. In \Cref{sec:kin_eq} we justify the use of the kinetic equilibrium distribution in the propagator. Briefly stated, gauge, scalar, and flavor-conserving Yukawa interactions drive the scalar propagators into equilibrium, while flavor-changing Yukawa interactions with out-of-equilibrium $N$ drive it out of equilibrium. We can then parametrize the correlations between the two mass eigenstates of $\eta$ with a chemical potential $\mu_{\eta 12}$, which is proportional to $\mu_N$, the chemical potential for $N$.

Following Ref. \cite{Garbrecht:2012pq}, we can integrate \cref{eq:kinetic} over the momentum $k$ where we separate the integrals for positive and negative $k^0$. Defining
\begin{equation}
n_{12}^\pm = 2 \int_0^{\pm \infty} \frac{d k^0}{2 \pi} \int \frac{d^3 k}{(2 \pi)^3} k^0 i D_{\eta 12} (k),
\end{equation}
we then obtain
\begin{equation}
\pm i (m_{\eta 1}^2 - m_{\eta 2}^2) n_{12}^\pm = - B_\eta^Y - B_\eta^{Y, \slashed{\text{fl}}} n_{12}^\pm - B_\eta^g (n_{12}^+ + n_{12}^-) - B_\eta^{\lambda, \text{even}} n_{12}^\pm - B_\eta^{\lambda, \text{odd}} n_{12}^\mp,
\label{eq:kinetic2}
\end{equation}
with the solution
\begin{subequations}
\begin{align}
q_{\eta 12} =& n_{12}^+ - n_{12}^- = \mathcal{R}_\eta 2 i B_\eta^Y, \\
\mathcal{R}_\eta =& \frac{m_{\eta 1}^2 - m_{\eta 2}^2}{(m_{\eta 1}^2 - m_{\eta 2}^2)^2 + (B_\eta^{Y, \slashed{\text{fl}}} + B_\eta^{\lambda, \text{even}} - B_\eta^{\lambda, \text{odd}})  (B_\eta^{Y, \slashed{\text{fl}}} + 2 B_\eta^g + B_\eta^{\lambda, \text{even}} + B_\eta^{\lambda, \text{odd}})}.
\end{align}
\end{subequations}
Here, $B_\eta^Y$ and $B_\eta^{Y, \slashed{\text{fl}}}$ are averaged rates for Yukawa-mediated flavor sensitive and flavor blind reactions, respectively, $B_\eta^{\lambda, \text{even}}$ and $B_\eta^{\lambda, \text{odd}}$ are the rates for scalar-mediated charge even and odd interactions, while $B_\eta^g$ is the averaged rate of (flavor blind) gauge processes. They are estimated in \Cref{sec:self_energies}. We can then relate the charge $q_{\eta 12}$ to the chemical potential $\mu_{\eta 12}$ with
\begin{equation}
q_{\eta 12} = 2 \int \frac{d^4 k}{(2 \pi)^4} k^0 i D_{\eta 12} (k) = \frac{\mu_{\eta 12} T^2}{3}.
\end{equation}

As for the vertex contribution to the $CP$-violating term, the source term is given by
\begin{equation}
	S_{\ell i}^{\rm v} = \int \frac{d^4 k}{2 \pi} \text{tr} [i \Sigma_\ell^{\text{v}, >} (k) i S_\ell^< (k) - i \Sigma_\ell^{\text{v}, <} (k) i S_\ell^> (k)],
\end{equation}
with \cite{Beneke:2010wd}
\begin{align}
\begin{split}
 i \Sigma_i^{\text{v}, ab} (k) =& - c d Y_i^{(a) *} Y_j^{(a)} Y_j^{(b) *} Y_i^{(b)} \int \frac{d^4 p}{(2 \pi)^4} \frac{d^4 q}{(2 \pi)^4} P_R i S_N^{ac} (-p) C [P_L i S_{\ell j}^{dc} (p+k+q) P_R]^t C^\dagger \\
	& \times i S_N^{db} (-q) P_L i \Delta_{\eta a}^{da} (-p-k) i \Delta_{\eta b}^{bc} (-q-k).
\end{split}
\end{align}
We present a detailed derivation of the vertex contribution in \cref{sec:vertex_contribution}.

The equilibration rates for $N$ and $\ell$ at tree-level are given by \cite{Garbrecht:2012qv,Garbrecht:2012pq}
\begin{align}
\mathcal{C}_N =& - \sum_i \left|Y_i^{(1)} \right|^2 \frac{a_R}{8 \pi M_N} z \frac{K_1 (z)}{K_2 (z)} \approx - \sum_i \left|Y_i^{(1)} \right|^2 \frac{a_R}{8 \pi M_N} z \equiv c_N z,
\label{eq:N_eq} \\
W_{\ell i} =& - \left|Y_i^{(1)} \right|^2 \frac{3 a_R}{8 \pi^3 M_N} z^3 K_1 (z) \approx - \left|Y_i^{(1)} \right|^2 3 \times 2^{-7/2} \pi^{-5/2} \frac{a_R}{M_N} z^{5/2} e^{-z} \equiv c_{W i} z^{5/2} e^{-z},
\label{eq:ell_eq}
\end{align}
while for the source terms we find
\begin{align}
\begin{split}
S_{\ell i}^\text{wf} =& \mathcal{R}_\eta \text{Im} [Y_i^{(1)} Y_j^{(1)*} Y_j^{(2)} Y_i^{(2)*}] \frac{3 a_R M_N z^5}{2^6 \pi^4} K_1 (z)\\
\approx& \mathcal{R}_\eta \text{Im} [Y_i^{(1)} Y_j^{(1)*} Y_j^{(2)} Y_i^{(2)*}] 3 \times 2^{-13/2} \pi^{-7/2} a_R M_N z^{5/2} e^{-z} \equiv c_{S i}^\text{wf} z^{5/2} e^{-z}.
\end{split}
\end{align}
and
\begin{align}
\begin{split}
	S_{\ell i}^\text{v} =& \text{Im} [Y_i^{(1)} Y_j^{(1)*} Y_j^{(2)} Y_i^{(2)*}] \frac{M_N^2}{m_2^2} \frac{a_R}{M_N} \frac{1}{2^8 \pi^2} \frac{(6 \sqrt{6} K_1 (\sqrt{6} z)/z + 4 K_0 (\sqrt{6} z))}{K_2 (z)} \\
	\approx& \text{Im} [Y_i^{(1)} Y_j^{(1)*} Y_j^{(2)} Y_i^{(2)*}] \frac{M_N^2}{m_2^2} \frac{a_R}{M_N} \pi^{-2} 2^{-25/4} 3^{-1/4} e^{(1-\sqrt{6})z} \equiv c_{Si}^\text{v} e^{(1-\sqrt{6})z}.
\end{split}
\end{align}
Note that, since $m_{\eta 2} \gg M_N, m_{\eta 1}$, the Yukawa couplings to $\eta_2$ do not enter the equilibration rates \cref{eq:N_eq,eq:ell_eq}. In addition, we have
\begin{equation}
Y_N^\text{eq} \approx \frac{2}{s} \int \frac{d^3 k}{(2 \pi)^3} e^{-k^0/T} = \pi^{-2} a_R^3 z^2 K_2 (z)/s \approx 2^{-1/2} \pi^{-3/2} a_R^3 z^{3/2} e^{-z}/s \equiv c_Y z^{3/2} e^{-z}.
\end{equation}
Since \cref{eq:kinetic_ell} is linear in $Y_{\ell i}$, we can decompose it into two equations
\begin{equation}
	\frac{d Y_{\ell i}^\text{wf,v}}{d z} = S_{\ell i}^\text{wf,v} (Y_N - Y_N^\text{eq}) + W_{\ell i} Y_{\ell i}^\text{wf,v},
\end{equation}
and add the two solutions to obtain the total yield. We can formally integrate the kinetic equations and obtain, for vanishing initial lepton asymmetry
\begin{equation}
Y_{\ell i} (z) = \int_{z_i}^z S_{\ell i} (z') (Y_N - Y_N^\text{eq}) (z') e^{\int_{z'}^z W_{\ell i} (z'') dz''} d z' = \int_{z_i}^z \frac{S_{\ell i} (z')}{\mathcal{C}_N (z')} \frac{d Y_N}{d z'} e^{\int_{z'}^z W_{\ell i} (z'') dz''} d z',
\end{equation}
which, in the strong washout regime, we can approximate as
\begin{equation}
Y_{\ell i} (z) = \int_{z_i}^z \frac{S_{\ell i} (z')}{\mathcal{C}_N (z')} \frac{d Y_N^\text{eq}}{d z'} e^{\int_{z'}^z W_{\ell i} (z'') dz''} d z'.
\end{equation}
Using Laplace's method, we can express the final asymmetries as \cite{Buchmuller:2004nz}
\begin{equation}
Y_{\ell i}^\text{wf} (z=\infty) = -\frac{c_{S i}^\text{wf}}{c_N} c_Y \left(z_{\text{fi,wf}} - \frac{3}{2} \right) z_{\text{fi,wf}}^2 \sqrt{ \frac{2 \pi e^{z_{\text{fi,wf}}}}{c_{Wi} (5/2 z_{\text{fi,wf}}^{3/2} - z_{\text{fi,wf}}^{5/2})}} e^{-\int_{z_{\text{fi,wf}}}^\infty c_{W i} z'^{5/2} e^{-z'} dz' - 2 z_{\text{fi,wf}}},
\label{eq:laplace}
\end{equation}
and
\begin{equation}
	Y_{\ell i}^\text{v} (z=\infty) = -\frac{c_{S i}^\text{v}}{c_N} c_Y \left(z_{\text{fi,v}} - \frac{3}{2} \right) z_{\text{fi,v}}^{-1/2} \sqrt{\frac{2 \pi e^{z_{\text{fi,v}}}}{c_W (5/2 z_{\text{fi,v}}^{3/2} - z_{\text{fi,v}}^{5/2})}} e^{-\int_{z_{fi}}^\infty c_{W i} z'^{5/2} e^{-z'} dz' - \sqrt{6} z_{\text{fi,v}}},
\end{equation}
with
\begin{equation}
z_{\text{fi,wf}} = - \frac{5}{2} W_{-1} \left(- \frac{2}{5} \times \left(\frac{2}{c_{Wi}} \right)^{2/5} \right),
\end{equation}
and
\begin{equation}
z_{\text{fi,v}} = - \frac{5}{2} W_{-1} \left(- \frac{2}{5} \times \frac{6^{1/5}}{c_{Wi}^{2/5}} \right),
\end{equation}
where $W_{-1}$ is the lower branch of the Lambert $W$ function.

The lepton asymmetry obtained is then transferred to the baryon sector through sphalerons. We can relate the final yield to the baryon asymmetry of the Universe through \cite{Harvey:1990qw,Chung:2008gv}
\begin{equation}
\eta_B = \frac{n_B}{n_\gamma} = 7.04 \times Y_{\ell i} \frac{24 + 4 N_\phi}{66 + 13 N_\phi},
\end{equation}
where $N_\phi$ is the number of Higgs doublets, in our case $N_\phi = 3$, and compare with the value obtained by the Planck collaboration \cite{Planck:2018vyg}
\begin{equation}
\eta_B = (6.143 \pm 0.190) \times 10^{-10}.
\end{equation}

\section{Dark Matter Production}
\label{sec:dark_matter}

\begin{figure}[h]
\centering
\subfloat[]{
\centering
\includegraphics[width=0.5\textwidth]{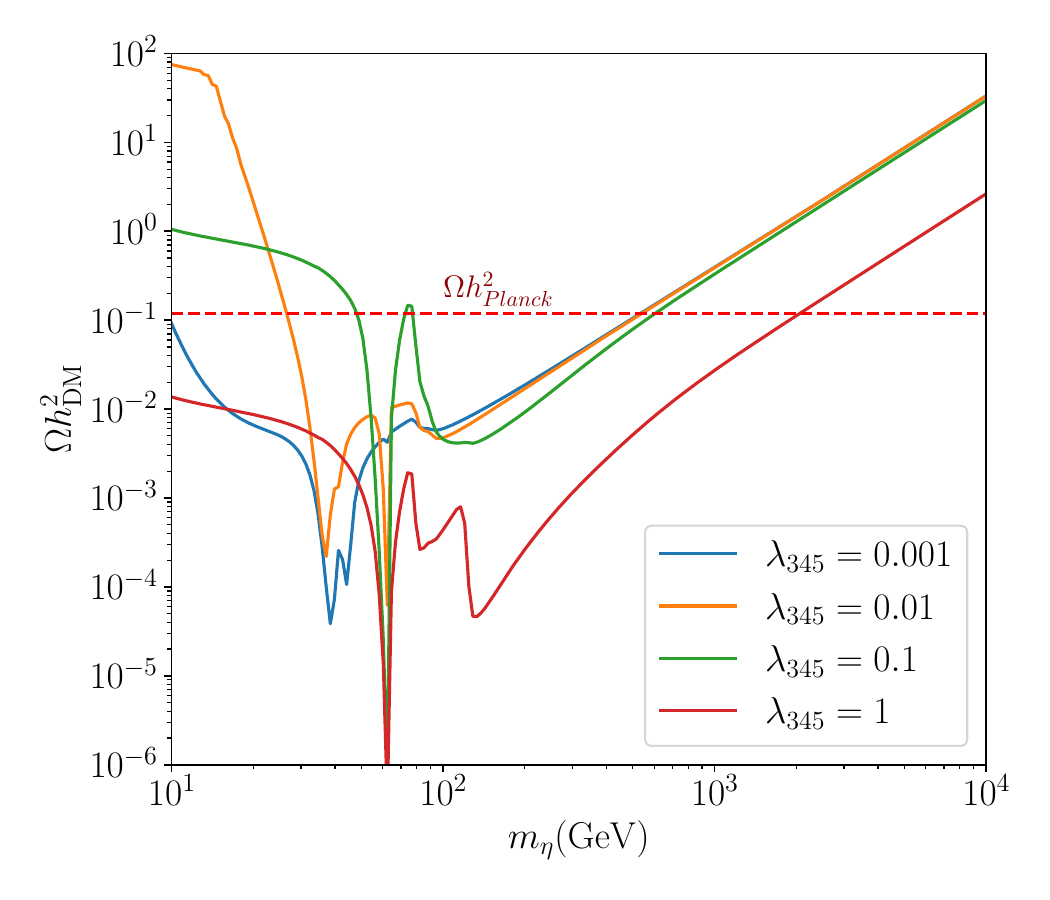}
\label{fig:dm_scan_a}} \hspace{-2em}
\subfloat[]{
\centering
\includegraphics[width=0.5\textwidth]{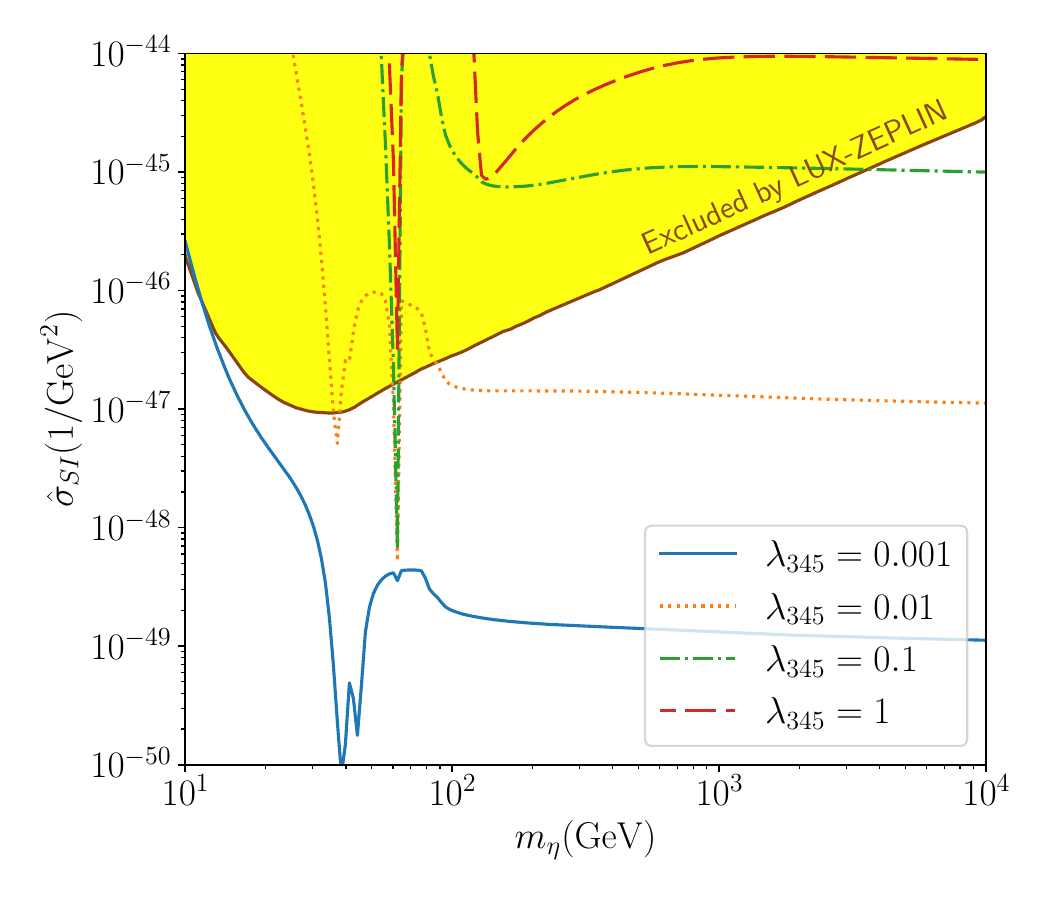}
\label{fig:dm_scan_b}}
\caption{Scan of dark matter relic abundance (a) and spin-independent cross section (b) in $m_\eta$ for different values of $\lambda$ compared with the Planck measurement \cite{Planck:2018vyg} and with LUX-ZEPLIN constraints \cite{LZ:2022lsv}, respectively.}
\label{fig:dm_scan}
\end{figure}

Given the mass relations we choose for this model, $\eta_1$, being the lightest particle in the $Z_2$-odd sector, is absolutely stable and therefore a natural dark matter candidate. This situation is similar to inert doublet models (IDM), which have been studied extensively in the literature \cite{Belyaev:2016lok, Branco:2011iw}. Being a scalar doublet, there is a wide range of processes that contribute to its annihilation rate, both via electroweak interactions and through the additional scalar couplings. We derive the Feynman rules for the model using FeynRules \cite{Alloul:2013bka} and compute the DM properties using micrOMEGAs \cite{Belanger:2013oya}. We choose $\lambda_5 < 0$ and set $\lambda_3 = \lambda_4 = - 0.4 \lambda_5$ as benchmark values. We then perform a scan of the DM relic abundance and cross section varying $m_{\eta 1}$ and compare with the Planck measurement $\Omega h_\text{DM}^2 = 0.120 \pm 0.001$ \cite{Planck:2018vyg} and with LUX-ZEPLIN constraints \cite{LZ:2022lsv}, respectively, as shown in \Cref{fig:dm_scan}. For the comparison with LUX-ZEPLIN data, we rescale the cross section as $\hat{\sigma} = \sigma \Omega_{\rm DM}/\Omega_{\rm DM}^{\text{Planck}}$, i.e. we assume the local dark matter density to coincide with the value of the cosmological average.

To avoid the overproduction of dark matter, a large $m_\eta$ needs to be compensated by large scalar couplings, which are constrained by direct detection. However, the direct detection bounds can be evaded if there are cancellations between $\lambda_3, \lambda_4$ and $\lambda_5$ in $\lambda_{345}$; in this case, the strongest constraint comes from the requirement of perturbativity. Combining both constraints, we find that $m_\eta$ must lie between around $\SI{375}{GeV}$ and $\SI{1000}{GeV}$ to be able to produce the correct relic abundance.

\section{Finding the Joint Parameter Space}
\label{sec:parameters}

The present model has the following free parameters: the Yukawa matrix $Y$, the masses $M_N, m_{\eta 1}, m_{\eta 2}$, and the scalar couplings $\lambda$. Since the Yukawa matrix, and therefore the neutrino mass matrix, has rank two, we can reduce the neutrino mass matrix to a $2 \times 2$ matrix, which, in the neutrino mass eigenbasis, is given by
\begin{equation}
m_\nu = Y \Lambda Y^T =
\begin{pmatrix}
m_2 & 0 \\
0 & m_3
\end{pmatrix},
\label{eq:diag_mass_matrix}
\end{equation}
where we define
\begin{equation}
\Lambda^{(a b)} = \frac{\lambda_5^{(ab)} v^2}{8 \pi^2} \frac{M_N}{m_{\eta_b}^2 - M_N^2} \left( \frac{m_{\eta_b}^2}{m_{\eta_a}^2 - m_{\eta_b}^2} \text{log} \left(\frac{m_{\eta_a}^2}{m_{\eta_b}^2} \right) - \frac{M_N^2}{m_{\eta_a}^2 - M_N^2} \text{log} \left( \frac{m_{\eta_a}^2}{M_N^2} \right) \right).
\label{eq:Lambda}
\end{equation}
For fixed Yukawa matrix $Y$ and particle masses, we can invert relation \cref{eq:diag_mass_matrix} to obtain
\begin{equation}
\lambda_5^{(ab)} = \left(Y^{-1}
\begin{pmatrix}
m_2 & 0 \\
0 & m_3
\end{pmatrix} \left( Y^T \right)^{-1}\right)^{(ab)}/\tilde{\Lambda}^{(ab)},
\label{eq:lambda}
\end{equation}
with $\tilde{\Lambda}^{(ab)} = \Lambda^{(ab)}/\lambda_5^{(ab)}$.

The lepton asymmetry is produced in two lepton flavors with opposite signs. We therefore need different washout rates $c_{W1} \neq c_{W2}$, otherwise the asymmetries would exactly cancel, and the final asymmetry is maximized in the strongly hierarchical case $c_{Wi} \gg c_{Wj}$. In addition to this, the mass splitting between the two neutral components is proportional to Re$\big(\lambda_5^{(11)} \big)$ per \cref{eq:mass_splitting}, which we also want to maximize, in order to avoid the direct detection via $Z$-boson exchange. We further choose Re$\big(\lambda_5^{(11)} \big)<0$. We find that setting $\phi \big(Y_1^{(1)} \big) = \pi/4, \phi \big(Y_2^{(1)} \big) = - \pi/4$ and $\phi \big(Y_1^{(2)} \big) = \phi \big(Y_2^{(2)} \big)$ arbitrary, we maximize the $CP$-violating phase and obtain $\phi \big(\lambda_5^{(11)} \big) = \pi$.

\begin{figure}[h!]
\includegraphics[width=\textwidth]{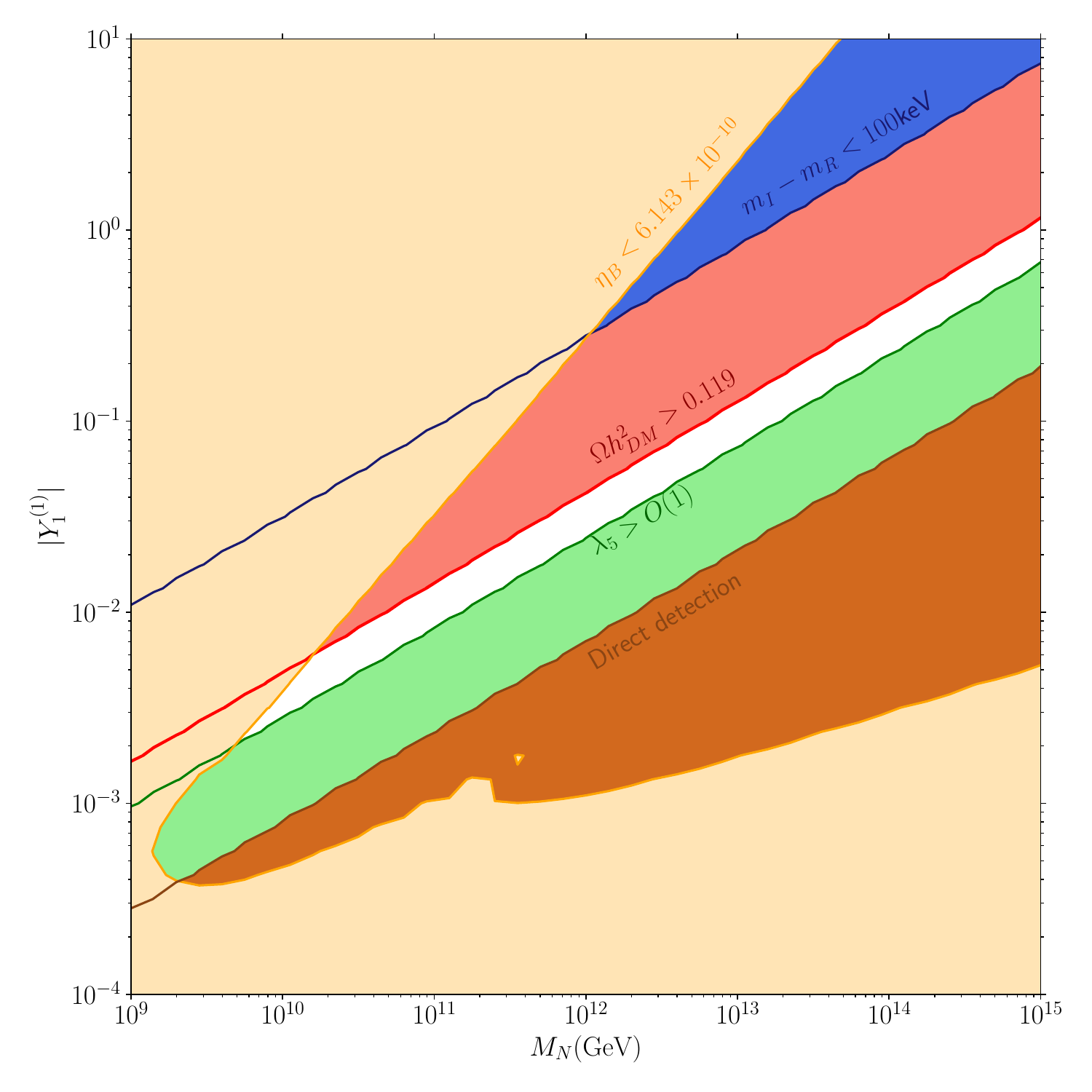}
\caption{Allowed region (white) for $M_N$ and $|Y_1^{(1)}|$ in the hierarchy $m_{\eta 2} > M_N > m_{\eta 1}$. We fix the ratios $|Y_2^{(1)}| = 0.1 |Y_1^{(1)}|$ as well as $|Y_1^{(2)}| = |Y_2^{(2)}| = 10^3 |Y_1^{(1)}|$ and $m_{\eta 2} = 10 M_N$, and set $m_{\eta 1} = \SI{750}{GeV}$. The orange region cannot account for the BAU, the blue line is where the mass splitting of the neutral scalars is too small to prevent upscattering via $Z$-boson exchange. The red line marks the constraint from dark matter overproduction, and the green line is where scalar interactions become larger than $O(1)$. The brown line is the direct detection bound from LUX-ZEPLIN.}
\label{fig:eta-decay}
\end{figure}

For scanning the parameter space, we fix $|Y_2^{(1)}| = 0.1 |Y_1^{(1)}|$ and $m_{\eta 1} = \SI{750}{GeV}$. Furthermore, since the impact of the Yukawa couplings to $\eta_2$ on the equilibration and washout rates are negligible, they are arbitrary. For illustrative purposes, we set $|Y_1^{(2)}| = |Y_2^{(2)}| = 10^3 |Y_1^{(1)}|$. Increasing this ratio can lead to leptogenesis scales down to $\SI{e7}{GeV}$, but even ratios down to $10$ give an allowed parameter region. The parameter scan with the relevant constraints is shown in \cref{fig:eta-decay}.

We find there is a region, albeit small, of values consistent with all bounds. Except for the baryon asymmetry, all bounds follow similar curves, which essentially depend on $\lambda_5$. From \Cref{eq:mass_matrix,eq:Lambda,eq:lambda}, we see that decreasing $|Y_1|$ and increasing $M_N$ leads to an increase in $\lambda_5$. Small values of $\lambda_5$ are constrained first because this would lead to DM overproduction (red region) and a small mass splitting between $m_R$ and $m_I$, making it susceptible to direct detection via neutral currents, while large values of $\lambda_5$ would violate perturbativity and would be inconsistent with direct detection bounds.

\section{Experimental Signatures}

As discussed in \cref{sec:constraints}, the main experimental constraints come from lepton flavor violation and direct detection experiments. Due to the very large mass of $N$, lepton flavor violating processes are strongly suppressed, and an improvement of many orders of magnitude in LFV precision measurements is required before a signal from our model is expected. On the other hand, we find that LUX-ZEPLIN places stringent bounds on the scalar couplings of the new doublets. While, as we have argued in \Cref{sec:dark_matter}, detection can be avoided in case of cancellations between $\lambda_3, \lambda_4$ and $\lambda_5$, large cancellations would require severe fine-tuning.

A third detection prospect is via missing transverse energy in collider experiments. For these searches, our model has essentially the same signatures as the inert Higgs doublet model, whose main detection channels for collider searches include: mono-jet production ($p p \to \eta_R \eta_R j$), mono-$Z$ production ($p p \to \eta_R \eta_R Z$), mono-Higgs production ($g g \to \eta_R \eta_R H$ and $q \bar{q} \to \eta_R \eta_I H$) and vector boson fusion ($p p \to \eta \eta j j$) for hadron colliders as well as $e^+ e^- \to \eta_R \eta_I$ and $e^+ e^- \to \eta^+ \eta^-$ for electron-positron colliders \cite{Belyaev:2016lok,Poulose:2016lvz,Avila:2019hhv,Belyaev:2018ext,Kalinowski:2018kdn,Datta:2016nfz,Ghosh:2021noq,Datta:2016nfz,Wan:2018eaz,Dutta:2017lny,Kalinowski:2018ylg,Kalinowski:2019cxe}. It was found that masses up to $\gtrsim \SI{300}{GeV}$ can be probed at the HL-LHC \cite{Datta:2016nfz,Wan:2018eaz,Dutta:2017lny}, which, while unable to explain all of the dark matter within our model, is also a viable scenario. This can be improved with higher center-of-mass energies, for instance in ILC or CLIC, which can potentially probe masses up to $\SI{1}{TeV}$ \cite{Kalinowski:2018ylg,Kalinowski:2019cxe}.

\section{Conclusion}

In this work we have investigated the simultaneous neutrino mass generation, DM production, and leptogenesis from a minimal realization of the scotogenic model, with two additional scalar doublets and a single Majorana fermion only, odd under a $\mathbb{Z}_2$ symmetry.

The leptogenesis mechanism we consider is via heavy Majorana fermion decay, where the decay asymmetry arises from the mixing and interferences between the two dark scalars. In our treatment of leptogenesis we have included an improved estimate of the scalar propagator width which strongly limits the resonant enhancement of the asymmetry. Due to the large width of the propagator, we find that in some regions of the parameter space vertex contributions to the $CP$-violating source term can become more relevant than the resonant wavefunction contributions. We also find that, on the one hand, large quartic couplings for the new scalars are needed to avoid the overproduction of dark matter, while on the other hand, direct detection experiments place stringent bounds on these quartic couplings, which can only be evaded if some cancellation between the scalar couplings occurs.

Nevertheless, performing a scan of $M_N$ and of the Yukawa couplings, for $\SI{500}{GeV}<m_\eta<\SI{1000}{GeV}$ we find a region of parameter space consistent with all constraints, which could explain the neutrino masses, account for the dark matter and the baryon asymmetry of the Universe simultaneously. Since there is a single Majorana fermion in the model, the $CP$-violating processes differ substantially from those in standard leptogenesis. In particular, these involve the interference of new particles that, while in the dark sector, are not Standard Model singlets. In the future, it may be interesting to pursue similar possibilities more broadly.

\section*{Acknowledgements}

E.W.'s work is funded by the Deutsche Forschungsgemeinschaft (DFG, German Research Foundation)–SFB 1258–283604770. We would also like to thank Carlos Tamarit for useful discussions.

\appendix

\section{Kinetic Equilibration of the Propagator}
\label{sec:kin_eq}

In order for \cref{eq:propagator} to hold, kinetic equilibrium has to be established for the scalar doublets $\eta$. Since $N$ is out of equilibrium, $\Pi^Y$ drives $i D_{12}$ also out of equilibrium. Since we do not make assumptions about the distribution of $N$, it is not immediately clear why $i D_{12}$ can still be taken to be of the form of a kinetic equilibrium distribution. This approximation, which is often used for calculations in different scenarios for baryogenesis, shall be justified in the present appendix. From the kinetic equations for the resummed scalar propagator we have \cite{Garbrecht:2011aw,Garny:2011hg}
\begin{equation}
2 k^0 \partial_t i D_{ij}^{<,>} + i(m_{\eta i}^2 - m_{\eta j}^2) iD_{ij}^{<,>} = - \frac{1}{2} ( i \Pi_{ik}^> i D_{kj}^< + i \Pi_{kj}^> i D_{ik}^< - i \Pi_{ik}^< i D_{kj}^> - i \Pi_{kj}^< i D_{ik}^> ).
\label{eq:kinetic_2}
\end{equation}

In the current discussion, we only include contributions to the self-energy from Yukawa and gauge interactions, $\Pi^Y$ and $\Pi^g$ respectively. For the moment, consider the case where $\Pi^Y = (M_1^2 - m_{\eta 2}^2) = 0$, so that the scalar particles are degenerate in their masses and interactions and that there are only gauge interactions. Looking for stationary solutions, \cref{eq:kinetic_2} then reduces to
\begin{equation}
i \Pi_{ik}^> i D_{kj}^< + i \Pi_{kj}^> i D_{ik}^< - i \Pi_{ik}^< i D_{kj}^> - i \Pi_{kj}^< i D_{ik}^> = 0.
\label{eq:equilibrium}
\end{equation}
This can be solved by assuming that the propagators follow a kinetic equilibrium distribution, writing, as in Ref. \cite{Beneke:2010dz},
\begin{subequations}
\begin{align}
i D_{ab}^< (k) =& 2 \pi \delta (k^2) [\theta (k^0) f_{ab}^\mu (k) + \theta (-k^0) (\mathbbm{1}_{ab} + \bar{f}_{ab}^\mu (k))], \\
i D_{ab}^> (k) =& 2 \pi \delta (k^2) [\theta (k^0) (\mathbbm{1}_{ab} + f_{ab}^\mu (k)) + \theta (-k^0) \bar{f}_{ab}^\mu (k)],
\end{align}
\label{eq:mixed_propagator}
\end{subequations}
with
\begin{equation}
f_{ab}^\mu (k) = \left( \frac{1}{e^{(|k| - \mu)/T} - 1} \right)_{ab}, \quad \bar{f}_{ab}^\mu (k) = \left( \frac{1}{e^{(|k| + \mu)/T} - 1}, \right)_{ab},
\end{equation}
where we have introduced a matrix of chemical potentials $\mu_{ab}$. With this we obtain a generalized Kubo-Matrin-Schwinger (KMS) relation for the propagators in kinetic equilibrium
\begin{equation}
D^>_{ab} (k) = \big( e^{(k^0 - \mu)/T} \big)_{ac} D^<_{cb} (k) = D^<_{ac} (k) \big( e^{(k^0 - \mu)/T} \big)_{cb}.
\end{equation}
To see that Eqs. (\ref{eq:mixed_propagator}) hold, note that the self-energy contribution from gauge interactions is given by
\begin{equation}
i \Pi_{ab}^{g, cd} (k) = g^2 \int \frac{d^4 k'}{(2 \pi)^4} \frac{d^4 k''}{(2 \pi)^4} (2 \pi)^4 \delta^4 (k - k' - k'') i D_{ab}^{cd} (k') k''^\mu k''^\nu i \Delta_{\mu \nu}^{cd} (k''),
\end{equation}
where $a,b$ are CTP indices, $c,d$ are flavor indices and $i \Delta_{\mu \nu}$ is the full gauge boson propagator. Since $i \Delta$ is in thermal equilibrium, it also observes the KMS relation. We then find that
\begin{equation}
i \Pi_{ab}^{g, >} (k) = \big( e^{(k^0 - \mu)/T} \big)_{ac} \Pi^{g,<}_{cb} (k) = \Pi^{g,<}_{ac} (k) \big( e^{(k^0 - \mu)/T} \big)_{cb},
\end{equation}
where the last equality follows from the fact that $\big( e^{(k^0 - \mu)/T} \big)$ and $D$ commute. With this we find
\begin{equation}
i \Pi_{ik}^> i D_{kj}^< = i \Pi^<_{il} \big( e^{(k^0 - \mu)/T} \big)_{lk} \big( e^{-(k^0 - \mu)/T} \big)_{km} i D_{mj}^> = i \Pi^<_{ik} i D_{kj}^>,
\end{equation}
and verify that \cref{eq:equilibrium} is indeed satisfied. To first order in the chemical potentials, we can approximate
\begin{equation}
f_{ab}^\mu (k) = \mathbbm{1}_{ab} \frac{1}{e^{|k|/T} - 1} + \frac{\mu_{ab}}{T} \frac{e^{|k|/T}}{(e^{|k|/T}-1)^2}.
\end{equation}

We now restrict the discussion to the components of the propagator accounting for mixing of the scalar flavors, i.e. $i D_{12}^< = i D_{12}^> = i D_{12}$. The kinetic equation for this part of the propagator is
\begin{multline}
i(m_{\eta 1}^2 - m_{\eta 2}^2) i D_{12} = - \frac{1}{2} ( (i \Pi_{12}^{Y,>} + i \Pi_{12}^{g,>}) (i D_{11}^< + i D_{22}^<) \\
+ (i \Pi_{11}^{Y,>} + i \Pi_{11}^{g,>} + i \Pi_{22}^{Y,>} + i \Pi_{22}^{g,>}) i D_{12} - < \leftrightarrow >).
\end{multline}
with the stationary solution
\begin{equation}
i D_{12} = \frac{1}{-i(m_{\eta 1}^2 - m_{\eta 2}^2) + i \Pi^\mathcal{A}} \left(- \frac{1}{2} (i \Pi_{12}^{Y,>} + i \Pi_{12}^{g,>}) (i D_{11}^< + i D_{22}^<) - < \leftrightarrow > \right).
\label{eq:stationary}
\end{equation}
While $\Pi^g$ contains $i D$, $\Pi^Y$ does not, since to leading order it only entails a fermion loop. Defining
\begin{equation}
A^{g, cd} (k, k') = g^2 \int \frac{d^4 k''}{(2 \pi)^4} (2 \pi)^4 \delta^4 (k - k' - k'') k''^\mu k''^\nu i \Delta_{\mu \nu}^{cd} (k''),
\end{equation}
we can rewrite the kinetic equation as
\begin{align}
\nonumber i(m_{\eta 1}^2 - m_{\eta 2}^2) i D_{12} (k) =& - \frac{1}{2} ( i \Pi_{12}^{Y,>} (k) (i D_{11}^< (k) + i D_{22}^< (k)) + (i \Pi_{11}^{Y,>} (k) + i \Pi_{22}^{Y,>} (k) ) i D_{12} (k) \\
\nonumber & + \int \frac{d^4 k'}{(2 \pi)^4} i A^{g, >} (k, k') [(i D_{11}^> (k') + i D_{22}^> (k')) i D_{12} (k) \\
& + i D_{12} (k') (i D_{11}^< (k) + i D_{22}^< (k))] - < \leftrightarrow >).
\label{eq:mixed_kinetic}
\end{align}
We now define $f (-k^0) = \bar{f} (|k^0|)$ and insert the form of the propagators given in Eqs. (\ref{eq:mixed_propagator}). We then find that \cref{eq:mixed_kinetic} takes the form
\begin{equation}
i(m_{\eta 1}^2 - m_{\eta 2}^2) f_{12} (k^0) = A_\text{const}^Y (k^0) + A_\text{diag}^Y (k^0) f_{12} (k^0) + A_\text{diag}^g (k^0) f_{12} (k^0) + \int \frac{d k'^0}{2 \pi} A_\text{nondiag}^g (k, k'^0) f_{12} (k'^0),
\label{eq:integrated}
\end{equation}
where we have introduced the short-hand notation
\begin{subequations}
\begin{align}
A_\text{const}^Y (k^0) =& - \frac{1}{2} i \Pi_{12}^{Y,>} (k) (f_{11} (k^0) + f_{22} (k^0)), \\
A_\text{diag}^Y (k^0) =&- \frac{1}{2} (i \Pi_{11}^{Y,>} (k) + i \Pi_{22}^{Y,>} (k)), \\
A_\text{diag}^g (k^0) =& \int \frac{d k'^4}{(2 \pi)^4} i A^{g, >} (k, k') [(i D_{11}^> (k') + i D_{22}^> (k')) - < \leftrightarrow >], \\
A_\text{nondiag}^g (k, k'^0) =& \int \frac{d k'^3}{(2 \pi)^3} 2 \pi \delta (k'^2) i A^{g, >} (k, k') [(f_{11} (k^0) + f_{22} (k^0)) - < \leftrightarrow >].
\end{align}
\end{subequations}
The subscripts here indicate the dependence on $f_{12}$, i.e. whether the terms on the right-hand side of \cref{eq:integrated} are independent of $f_{12}$ and if they are dependent, whether they are diagonal or generally nondiagonal in $k^0$.

When we discretize the momentum $k^0$, we obtain
\begin{equation}
i(m_{\eta 1}^2 - m_{\eta 2}^2) f_{12} (k_i) = A_\text{const}^Y (k_i) + (A_\text{diag}^Y (k_i) + A_\text{diag}^g (k_i)) f_{12} (k_i) + \sum_j \frac{\Delta k}{2 \pi} A_\text{nondiag}^g (k_i, k_j) f_{12} (k_j),
\label{eq:discretized}
\end{equation}
which we can interpret as a matrix equation
\begin{equation}
R_{ij} f_{12} (k_j) = y v_i,
\label{eq:matrix_eq}
\end{equation}
where $y$ is a new expansion parameter which we tag to all quantities driving the distribution out of equilibrium.

\paragraph{Quasidegenerate case} Here, the mass splitting $m_{\eta 1}^2 - m_{\eta 2}^2$ as well as the interactions mediated by $Y$ are small compared to the gauge interactions mediated by $g$. The latter thus have time to establish kinetic equilibrium also in the off-diagonal correlations. We thus decompose $R$ as
\begin{align}
v_i =& A_\text{const}^Y (k_i), \\
R_{ij} =& R_{ij}^g + y R_{ij}^Y = - A_\text{diag}^g (k_i) \delta_{ij} -\frac{\Delta k}{2 \pi} A_\text{nondiag}^g (k_i, k_j) - y (A_\text{diag}^Y (k_i) - i(m_{\eta 1}^2-m_{\eta 2}^2)) \delta_{ij},
\end{align}
Clearly, the solution to \cref{eq:matrix_eq} is
\begin{equation}
f_{12} (k_i) = y R_{ij}^{-1} v_j.
\label{eq:solution}
\end{equation}
We also recover the pure gauge scenario when we send $y \to 0$, in which case \cref{eq:matrix_eq} becomes
\begin{equation}
R_{ij}^g f_{12} (k_j) = 0,
\end{equation}
whose solution is precisely the chemical equilibrium distribution $f_{12}^\mu$, as we have seen before. From this equation we also recognize that $R_{ij}^g$ is singular. Then, following Ref. \cite{Avrachenkov}, we can regard $R_{ij}$ as a matrix function in $y$ and expand $R_{ij}^{-1} (y)$ as a Laurent series in $y$
\begin{equation}
M^{-1} (y) = \frac{1}{y^s} (X_0 + y X_1 + \dots),
\end{equation}
where $s$ is the order of the pole at $y=0$. We expect the kernel of $T^g$ to be one-dimensional, and if it is not simultaneously the kernel of $R^Y$, using the method introduced in Ref. \cite{Sain1969}, we find that $R^{-1}$ has a simple pole at the origin, and from the condition
\begin{equation}
R R^{-1} = \mathbbm{1},
\end{equation}
we obtain the fundamental equations
\begin{subequations}
\begin{align}
R^g X_0 &= 0,
\label{eq:fundamental_zero} \\
R^g X_1 + R^Y X_0 &= \mathbbm{1}, \\
R^g X_2 + R^Y X_1 &= 0, \\
& \nonumber \vdots
\end{align}
\end{subequations}
With these equations, we can find $X_0, X_1 \dots$ and write our problem as
\begin{equation}
f_{12} (k) = X_0 v^Y + y X_1 v^Y + \mathcal{O} (y^2).
\end{equation}
To zeroth order in $y$, the solution \cref{eq:solution} is given by $X_0 v^Y$, but from \cref{eq:fundamental_zero}, we know that $M^g X_0 v^Y = 0$, which means that $X_0 v^Y$ lies in the kernel of $M^g$. Since the kernel of $M^g$ is one-dimensional, this implies that $X_0 v^Y$ is proportional to $f_{12}^\mu (k)$, and therefore, if the gauge interactions are much stronger than the Yukawa interactions and the squared mass difference of the $\eta$, $f_{12}$ can indeed be approximated by $f_{12}^\mu$, the kinetic equilibrium distribution.

\paragraph{Nondegenerate case} So far we have treated the mass splitting as a perturbation compared to the gauge self-energies. We now want to treat the case where the mass splitting dominates the kinetic equation. The gauge interactions then do not have time to impose kinetic equilibrium on the off-diagonal correlations induced by the out-of-equilibrium Majorana fermion. In this case, we can rewrite \cref{eq:matrix_eq} as
\begin{equation}
	(i \Delta M_\eta^2 \mathbbm{1} - M^{g + Y})_{ij} f_{12} (k_j) = v_i,
\label{eq:matrix_2}
\end{equation}
where
\begin{align}
v_i =& A_\text{const}^Y (k_i), \\
R_{ij}^{g + Y} =& A_\text{diag}^g (k_i) \delta_{ij} + \frac{\Delta k}{2 \pi} A_\text{nondiag}^g (k_i, k_j) + A_\text{diag}^Y (k_i) \delta_{ij}.
\end{align}
We can divide \cref{eq:matrix_2} by $i \Delta M_\eta^2$ and obtain
\begin{equation}
	\left( \mathbbm{1} - \frac{M^{g + Y}}{i \Delta M_\eta^2} \right)_{ij} f_{12} (k_j) = \frac{v_i}{i \Delta M_\eta^2},
\end{equation}
with the solution
\begin{equation}
	f_{12} (k_i) = \left( \mathbbm{1} - \frac{M^{g + Y}}{i \Delta M_\eta^2} \right)_{ij}^{-1} \frac{v_j}{i \Delta M_\eta^2}.
\end{equation}
Since we assume $|i \Delta M_\eta^2| \gg |M^{g + Y}|$, we can write $(\mathbbm{1} - M^{g + Y}/(i \Delta M_\eta^2))^{-1}$ as a Neumann series
\begin{equation}
	\left(\mathbbm{1} - \frac{M^{g + Y}}{i \Delta M_\eta^2} \right)^{-1} = \sum_k \left( \frac{M^{g + Y}}{i \Delta M_\eta^2} \right)^k.
\end{equation}
Since higher order terms are suppressed by powers of $1/\Delta M_\eta^2$, we can keep only the leading order term, which gives us
\begin{equation}
	f_{12} (k_i) = \frac{v_i}{i \Delta M_\eta^2}.
\end{equation}

Going back to \cref{eq:stationary}, this means we can approximate
\begin{equation}
	i D_{12} \approx \frac{1}{-i(m_{\eta 1}^2 - m_{\eta 2}^2)} \left(- \frac{1}{2} i \Pi_{12}^{Y,>} (i D_{11}^< + i D_{22}^<) - < \leftrightarrow > \right).
\label{eq:stationary_2}
\end{equation}
If we parametrize the deviation of $N$ from equilibrium with a pseudo-chemical potential $\mu_N$, we find again a KMS relation for $\Pi_{12}^Y$:
\begin{equation}
	i \Pi_{12}^{Y, >} (k) = e^{(k^0 - \mu_N)/T} i \Pi_{12}^{Y, <} (k),
\end{equation}
from which we can rewrite \cref{eq:stationary_2} as
 \begin{equation}
	i D_{12} (k) = \frac{1}{-i(m_{\eta 1}^2 - m_{\eta 2}^2)} 2 \Pi_{12}^{Y,\mathcal{A}} (k) \Delta_{\eta 1}^\mathcal{A} (k) \frac{ \text{sign} (k^0)  (1 - e^{- \mu_N/T}) e^{k^0/T}}{(e^{(k^0 - \mu_N)/T} - 1)(e^{k^0/T} - 1)},
\end{equation}
where we have omitted the contribution from $\Delta_{\eta 2}$ since we assume $\eta_2$ to be much heavier than $\eta_1$, and can therefore neglect its on-shell contribution. To first order in the chemical potential we can write
\begin{equation}
	i D_{12} (k) = 2 \pi \delta(k^2 - m_{\eta 1}^2) \frac{1}{-i(m_{\eta 1}^2 - m_{\eta 2}^2)} \Pi_{12}^{Y,\mathcal{A}} (k) \frac{\mu_N}{T} \frac{ \text{sign} (k^0)  e^{|k^0|/T}}{(e^{|k^0|/T} - 1)^2}.
\end{equation}
In the regime of interest, $m_{\eta 1} \gg T$, the dominant contributions arise for $\mathbf{k} \ll T$ and $|k^0| - m_{\eta 1} \ll T$, so that we can neglect the $k$ dependence of $\Pi_{12}^{Y,\mathcal{A}} (k)$ and rewrite this as 
\begin{equation}
	i D_{12} (k) = 2 \pi \delta(k^2 - m_{\eta 1}^2) \frac{\mu_{\eta 12}}{T} \frac{ \text{sign} (k^0)  e^{|k^0|/T}}{(e^{|k^0|/T} - 1)^2},
\end{equation}
where we have introduced a new chemical potential $\mu_{\eta 12}$, chosen in such a way that both definitions produce the same charges. So we can again apply \cref{eq:propagator}. Note that while the approximation does not apply in the exponential tail, it holds for the relevant momentum range.

With this we have shown that we can parametrize $f_{12}$ with a chemical potential both in the case where the kinetic equation (\ref{eq:kinetic_2}) is dominated by gauge (or other) interactions driving it to kinetic equilibrium as well as when it is dominated by the mass splitting.

\section{Spectral Self-Energies}
\label{sec:self_energies}

To determine the width of the mixed scalar propagator from the kinetic equation (\ref{eq:kinetic}), we need to compute the spectral self-energies for the fields. The two main contributions come from the Yukawa and the gauge interactions. In \cref{eq:kinetic2} we have introduced the averaged rates, which are defined as
\begin{align}
	B_\eta^Y &= \pm \int_0^{\pm \infty} \frac{d k^0}{2 \pi} \int \frac{d^3 k}{(2 \pi)^3} k^0 i \Pi_{12}^{Y, >} (k) (i \Delta_{\eta 11}^< (k) + i \Delta_{\eta 22}^< (k)) - < \leftrightarrow >, \\
	B_\eta^{Y, \slashed{\text{fl}}} &= \pm \frac{1}{n_{12}^\pm} \int_0^{\pm \infty} \frac{d k^0}{2 \pi} \int \frac{d^3 k}{(2 \pi)^3} 2 k^0 (\Pi_{11}^{Y, \mathcal{A}} (k) + \Pi_{22}^{Y, \mathcal{A}} (k)) i D_{\eta 12} (k), \\
	B_\eta^g &= \pm \frac{1}{n_{12}^\pm} \int_0^{\pm \infty} \frac{d k^0}{2 \pi} \int \frac{d^3 k}{(2 \pi)^3} 2 k^0 (\Pi_{11}^{g, \mathcal{A}} (k) + \Pi_{22}^{g, \mathcal{A}} (k)) i D_{\eta 12} (k), \\
	B_\eta^{\lambda, \text{even}} n_{12}^\pm + B_\eta^{\lambda, \text{odd}} n_{12}^\mp &= \pm \int_0^{\pm \infty} \frac{d k^0}{2 \pi} \int \frac{d^3 k}{(2 \pi)^3} k^0 \sum_k (i \Pi_{12}^> i \Delta_{\eta kk}^< + i \Pi_{kk}^> i D_{\eta 12} - < \leftrightarrow >).
\end{align}
Using CTP methods and assuming $M_N \gg T$, we find \cite{Garbrecht:2012pq}
\begin{equation}
	B_\eta^Y = [Y^\dagger Y]_{12} \frac{M_N^4 \mu_N}{32 \pi^3} K_2 (M_N/T),
\end{equation}
where $\mu_N$ is a chemical potential we introduce to parametrize the deviation of $N$ from equilibrium. The Yukawa spectral self-energy can be computed at first order, giving
\begin{equation}
	\Pi_{ab}^{Y, \mathcal{A}} (k) = - \text{sign} (k^0) \frac{[Y^\dagger Y]_{ab}}{16 \pi k^0} T M_N^2 e^{-M_N^2/(4 |k^0| T)},
\end{equation}
where we have approximated the distribution of the Majorana neutrinos $N$ as nonrelativistic. We then find
\begin{equation}
	B_\eta^{Y, \slashed{\text{fl}}} = \sum_j (Y_j^{(1), 2} + Y_j^{(2), 2})  \frac{3}{32 \pi^3} \frac{M_N^4}{T^2} K_2 (M_N/T).
\end{equation}

Defining the reduced cross section for a two-body scattering as
\begin{equation}
	\hat{\sigma} (s) = \frac{2 \lambda (s, m_1, m_2)}{s} \sigma (s),
\end{equation}
where $\sigma$ is the usual cross section and $\lambda$ is the function
\begin{equation}
	\lambda(s, m_1, m_2) = \sqrt{(s - (m_1 + m_2)^2)(s - (m_1 - m_2)^2)},
\end{equation}
one can compute the reaction density as
\begin{equation}
	\gamma = \frac{T}{64 \pi^4} \int_{(m_1 + m_2)^2}^\infty ds \hat{\sigma} (s) \sqrt{s} K_1 \left( \frac{\sqrt{s}}{T} \right).
\end{equation}

We present here a detailed derivation of $B^\lambda$, which, to the best of our knowledge, is computed for the first time. The relevant contribution to the self-energy comes from the sunset diagram and is given by
\begin{equation}
	i \Pi^{\lambda ab} (k) = \sum \lambda^\dagger \lambda \int \frac{d^4 p}{(2 \pi)^4} \frac{d^4 p'}{(2 \pi)^4} i \Delta_\phi^{ab} (p) i \Delta_{\phi/\eta}^{ab} (p') i \Delta_{\eta/\phi}^{ba} (p+p'-k),
\end{equation} 
where $\lambda^\dagger \lambda$ stands for the coupling structure and the sum is over the field configurations running in the loop. We also choose a signature where all temporal momenta have the same sign (since the expression is symmetric under exchange of the momenta, this can always be done). The collision term is then
\begin{align}
\begin{split}
	\mathcal{C} =& \sum_k (i \Pi_{12}^> i \Delta_{kk}^< + i \Pi_{kk}^> i D_{12} - i \Pi_{12}^< i \Delta_{kk}^> - i \Pi_{kk}^< i D_{12}) \\
	=& \sum_{k, j \neq \ell} \int \frac{d^4 p}{(2 \pi)^4} \frac{d^4 p'}{(2 \pi)^4} \frac{d^4 k'}{(2 \pi)^4} (2 \pi)^4 \delta^4 (k+k'-p-p') \\
	\times & \bigg\{i \Delta_\phi^> (p) \Big[\lambda_5^{1 j} \lambda_5^{\ell 2 *} i \Delta_\phi^> (p') i D_{\eta j \ell} (k') + \big(\lambda_3^{1j} + \lambda_4^{j1} + \text{h.c.} \big)\big(\lambda_3^{\ell 2 *} + \lambda_4^{2 \ell *} + \text{h.c.} \big) i D_{\eta j \ell} (p') i \Delta_\phi^< (k')\Big] i \Delta_{\eta kk}^< (k) \\
	+&  i \Delta_\phi^> (p) \Big[\lambda_5^{k \ell} \lambda_5^{\ell k*} i \Delta_\phi^> (p') i \Delta_{\eta \ell \ell}^< (k') + \big(\lambda_3^{k \ell} + \lambda_4^{\ell k} + \text{h.c.} \big) \big(\lambda_3^{\ell k *} + \lambda_4^{k \ell *} + \text{h.c.} \big) i \Delta_{\eta \ell \ell}^> (p') i \Delta_\phi^< (k')\Big] i D_{\eta 12} (k) \\
	- &i \Delta_\phi^< (p) \Big[\lambda_5^{1j} \lambda_5^{\ell 2 *}i \Delta_\phi^< (p') i D_{\eta j \ell} (k') + \big(\lambda_3^{1j} + \lambda_4^{j1} + \text{h.c.} \big) \big(\lambda_3^{\ell 2 *} + \lambda_4^{2 \ell *} + \text{h.c.} \big) i D_{\eta j \ell} (p') i \Delta_\phi^> (k')\Big] i \Delta_{\eta kk}^> (k) \\
	-& i \Delta_\phi^< (p) \Big[\lambda_5^{k \ell} \lambda_5^{\ell k*} i \Delta_\phi^< (p') i \Delta_{\eta \ell \ell}^> (k') + \big(\lambda_3^{k \ell} + \lambda_4^{\ell k} + \text{h.c.} \big)\big(\lambda_3^{\ell k *} + \lambda_4^{k \ell *} + \text{h.c.} \big) i \Delta_{\eta \ell \ell}^< (p') i \Delta_\phi^> (k')\Big] i D_{\eta 12} (k)\bigg\}.
\end{split}
\end{align}
While $D_{\eta 12}$ also contributes wherever $\Delta_{\eta \ell \ell}$ appears, this would be a higher order correction, which we discard as we only consider linear terms in the off-diagonal correlations of the particles $\eta_{1,2}$. Similarly, $\Delta_{\eta \ell \ell}$ also should contribute where $D_{\eta j \ell}$ appears, but since we assume diagonal propagators to be in equilibrium, this would give a purely equilibrium, and therefore vanishing, contribution. Since $m_{\eta 2} \gg T$, we can also neglect $\Delta_{\eta 22}$ contributions, since they are strongly Maxwell-suppressed. Lastly, we see that not only $D_{\eta 12}$ appears in the collision term but also $D_{\eta 21}$, implying that there is some mixing between $n_{12}$ and $n_{21}$. We also neglect this term to avoid complicating the problem even further. The full reaction rate is obtained by integrating the collision term over $k$, and, in spite of the complicated flavor structure, we can approximate all processes as having the same $2 \leftrightarrow 2$ kinematics. Paying particular attention to the signs of the contributions, this gives us
\begin{align}
\begin{split}
	\Gamma^{\lambda, \pm} =& \int_0^{\pm \infty} \frac{dk^0}{2 \pi} \int \frac{d^3 k}{(2 \pi)^3} \mathcal{C} (k) \\
	=& \frac{\Gamma}{2} [(\lambda_5^{11} \lambda_5^{22 *} - (\lambda_3^{11} + \lambda_4^{11} + h.c.)(\lambda_3^{22 *} + \lambda_4^{22 *} + h.c.)) (n_{12}^\pm - 2 n_{12}^\mp) \\
	& + 3 \textstyle \sum_{k} (\lambda_5^{k 1} \lambda_5^{1 k *} + (\lambda_3^{k 1} + \lambda_4^{1 k} + h.c.)(\lambda_3^{1 k *} + \lambda_4^{k 1 *} + h.c.)) n_{12}^\pm], \\
	\equiv & \Gamma^{\lambda, \text{even}} n_{12}^\pm + \Gamma^{\lambda, \text{odd}} n_{12}^\mp,
\end{split}
\end{align}
where we divide by two to average over the $SU(2)$ degrees of freedom, and
\begin{align}
\begin{split}
	\Gamma =& \frac{1}{n^\text{eq}} \int \frac{d^3 k}{(2 \pi)^3 2 |k^0|} \frac{d^3 k'}{(2 \pi)^3 2 |k'^0|} \frac{d^3 p}{(2 \pi)^3 2 |p^0|} \frac{d^3 p'}{(2 \pi)^3 2 |p'^0|} (2 \pi)^4 \delta^4 (k+k'-p-p') f (k) f (k')\\
	=& \frac{T}{16 \pi^2},
\end{split}
\end{align}
in Boltzmann approximation. Then, with the approximation that all particles have the same reaction rate regardless of momentum, we can write \cite{Garbrecht:2012pq}
\begin{equation}
	B n_{12}^\pm = \int \frac{d^3 k}{(2 \pi)^3} |k| \Gamma f_{12} (k) = n_{12}^\pm \frac{36}{\pi^2} \Gamma T.
\end{equation}

As for gauge interactions, since we are in a regime where the gauge bosons are massless, the first-order self-energy corresponding to $1 \leftrightarrow 2$ processes vanishes. We therefore need to go to second order, which corresponds to two-by-two scatterings. The only relevant scatterings are pair creation and annihilation since they are the only ones that can change particle number. We use FeynArts \cite{Hahn:2000kx} to generate the relevant diagrams and FeynCalc \cite{Shtabovenko:2020gxv,Shtabovenko:2016sxi,Mertig:1990an} to obtain the corresponding amplitudes. As opposed to earlier estimates from Ref. \cite{Garbrecht:2012pq}, IR divergences in $t$ and $u$-channel cancel,  and these contributions can be directly accounted for. The total reduced cross section we find is
\begin{equation}
\hat{\sigma} (s) = \frac{1}{12} (327 g_1^4 + 42 g_1^2 g_2^2 + 169 g_2^4),
\end{equation}
from which we obtain the reaction density
\begin{equation}
\gamma^g = \frac{T}{64 \pi^4} \int_0^\infty ds \sqrt{s} K_1 \left( \frac{\sqrt{s}}{T} \right) \hat{\sigma} (s) = \frac{T^4}{192 \pi^4} (327 g_1^4 + 42 g_1^2 g_2^2 + 169 g_2^4).
\end{equation}
From this, following Ref. \cite{Garbrecht:2012pq}, using $g_2 = 0.6$ and $g_1 = 0.4$ we find
\begin{equation}
B_\eta^g = \frac{36}{\pi^2} \Gamma^g T = \frac{36}{\pi^2} \frac{\gamma^g}{3/(2 \pi)^2 \zeta (3) T^2} = 1.4 \times 10^{-3} T^2,
\end{equation}
which is one order of magnitude larger than the expression in Ref. \cite{Garbrecht:2012pq}.

\section{Vertex Contribution to the Source Term}
\label{sec:vertex_contribution}

The vertex contribution to the $CP$-source is given by
\begin{align}
\begin{split}
	S_{\ell i}^\text{v} =& \int \frac{d^4 k}{(2 \pi)^4} \text{tr} [i \Sigma_\ell^> (k) i S_\ell^< (k) - i \Sigma_\ell^< (k) i S_\ell^> (k)] \\
	=& \int \frac{d^4 k}{(2 \pi)^4} \frac{d^4 p}{(2 \pi)^4} \frac{d^4 q}{(2 \pi)^4} \text{tr} [ - Y_i^{(a) *} Y_j^{(a)} Y_j^{(b) *} Y_i^{(b)} \{ i \delta S_N (-p) C[i S_{\ell j}^T (p+k+q) i \Delta_{\eta b}^T (-q - k) \\
	&- i S_{\ell j}^< (p + k + q) i \Delta_{\eta b}^< (-q - k)]^t C^\dagger i S_N^T (-q) i \Delta_{\eta a}^< (-p - k) \\
	& + i S_N^{\bar{T}} (-p) C[i S_{\ell j}^{\bar{T}} (p+k+q) i \Delta_{\eta a}^{\bar{T}} (-p-k) - i S_{\ell j}^< (p + k + q) i \Delta_{\eta a}^< (-p - k)]^t C^\dagger i \Delta_{\eta b}^< (-q - k) \\
	& i \delta S_N (-q) \} i S_{\ell i}^< (k)] - (+ \leftrightarrow -).
\end{split}
\end{align}
Dropping on-shell $\eta_2$ terms, we have
\begin{align}
\begin{split}
	S_{\ell i}^\text{v} =& \int \frac{d k_0}{2 \pi} \frac{d^4 p}{(2 \pi)^4} \frac{d^4 q}{(2 \pi)^4}\\
	& \text{tr} [\{- Y_i^{(1) *} Y_j^{(1)} Y_j^{(2) *} Y_i^{(2)} i \delta S_N (-p) C[i S_{\ell j}^T (p+k+q)]^t C^\dagger  i \Delta_{\eta 2}^T (-q - k) i S_N^T (-q) i \Delta_{\eta 1}^< (-p - k) \\
	&+ Y_i^{(1)} Y_j^{(1) *} Y_j^{(2)} Y_i^{(2) *} i S_N^{\bar{T}} (-p) C[i S_{\ell j}^{\bar{T}} (p+k+q)]^t C^\dagger i \Delta_{\eta 2}^{T} (-p-k) i \Delta_{\eta 1}^< (-q - k) i \delta S_N (-q)\} i S_{\ell i}^< (k) \\
	&+ \{-Y_i^{(1) *} Y_j^{(1)} Y_j^{(2) *} Y_i^{(2)} i \delta S_N (-p) C[i S_{\ell j}^{\bar{T}} (p+k+q)]^t C^\dagger i \Delta_{\eta 2}^{T} (-q - k) i S_N^{\bar{T}} (-q) i \Delta_{\eta 1}^> (-p - k) \\
	&+ Y_i^{(1)} Y_j^{(1) *} Y_j^{(2)} Y_i^{(2) *} i S_N^T (-p) C[i S_{\ell j}^T (p+k+q)]^t C^\dagger i \Delta_{\eta 2}^T (-p-k) i \Delta_{\eta 1}^> (-q - k) i \delta S_N (-q)\} i S_{\ell i}^> (k)],
\end{split}
\end{align}
where we have used $i \Delta_{\eta 2}^T = - i \Delta_{\eta 2}^{\bar{T}}$ for off-shell $\eta_2$.

As was argued in Ref. \cite{Beneke:2010wd}, we only need to keep the absorptive parts of $i S_{\ell}^{T, \bar{T}}$ and $i S_N^{T, \bar{T}}$, since the dispersive parts cancel upon integration. We then have
\begin{align*}
	S_{\ell i}^\text{v} =& - 4 \text{Im} [Y_i^{(1)} Y_j^{(1)*} Y_j^{(2)} Y_i^{(2)*}] \int \frac{d^3 p}{(2 \pi)^3 2 p^0} \frac{d^3 k}{(2 \pi)^3 2 k^0} \frac{d^3 p'}{(2 \pi)^3 2 p'^0} \frac{d^3 q}{(2 \pi)^3 2 q^0} \frac{d^3 q'}{(2 \pi)^3 2 q'^0} \delta^4 (q-q'-p') \delta^4 (p - k - q') \\
	&(k \cdot p')  \delta f_N (p) f_{\ell j} (p') f_N (q) \frac{M_N^2}{(q+k)^2 - m_{\eta 2}^2} (1 + f_{\eta 1} (q') - f_{\ell i} (k)).
\end{align*}
We can use the spatial delta functions to carry out the $p'$ and the $k$ integrals. Neglecting quantum statistical factors, we then have
\begin{align*}
	S_{\ell i}^\text{v} =& - 4 \text{Im} [Y_i^{(1)} Y_j^{(1)*} Y_j^{(2)} Y_i^{(2)*}] \int \frac{d^3 p}{(2 \pi)^3 2 p^0} \frac{d^3 q}{(2 \pi)^3 2 q^0} \frac{d^3 q'}{(2 \pi)^3 2 q'^0} \frac{1}{2 k^0} \frac{1}{2 p'^0}  (2 \pi)^2 \delta (q^0-q'^0-p'^0) \delta (p^0 - k^0 - q'^0) \\
	& ((q-q') \cdot (p-q')) \delta f_N (p) f_{\ell j} (q-q') f_N (q) \frac{M_N^2 }{(q+k)^2 - m_{\eta 2}^2}.
\end{align*}
We can use spherical coordinates to express $p$ and $q$ with respect to $q'$ through the angles $\theta_{p, q}, \varphi_{p, q}$ and use the delta functions to do the $\theta_{p, q}$ integrals. Approximating $M_N^2/((q+k)^2 - m_{\eta 2}^2) \approx - M_N^2/m_{\eta 2}^2$, we have
\begin{align*}
	S_{\ell i}^\text{v} =& 4 \text{Im} [Y_i^{(1)} Y_j^{(1)*} Y_j^{(2)} Y_i^{(2)*}] \int \frac{d^3 q'}{(2 \pi)^3 2 q'^0} \int_{|q' - M_N^2/(4 q')|}^\infty \frac{dp}{2 \pi} \frac{d \varphi_p p^2}{(2 \pi) 2 p^0} \int_{|q' - M_N^2/(4 q')|}^\infty \frac{d q}{2 \pi} \frac{d \varphi_q q^2}{(2 \pi) 2 q^0} \\
	&\frac{1}{2 p q'} \frac{1}{2 q q'} (2 M_N^2 (q-q') \cdot (p-q')) \delta f_N (p) f_{\ell j} (q-q') f_N (q) \frac{M_N^2}{m_{\eta 2}^2}.
\end{align*}
We express
\begin{align*}
	(q-q') \cdot (p-q') =& - \frac{1}{4} p q \sqrt{-\frac{M_N^2 (4 q' (q' - p^0) + M_N^2)}{p^2 q'^2}} \sqrt{-\frac{M_N^2 (4 q' (q' - q^0) + M_N^2)}{q^2 q'^2}} \text{cos} (\varphi_p - \varphi_q) \\
	&- \frac{(M_N^2 - 4 q'^2)(M_N^2 - 2 q' (p^0 + q^0))}{4 q'^2}.
\end{align*}
The first term vanishes when performing the $\varphi$ integrals, and we are left with
\begin{align*}
	S_{\ell i}^\text{v} =& 4 \text{Im} [Y_i^{(1)} Y_j^{(1)*} Y_j^{(2)} Y_i^{(2)*}] \frac{M_N^2}{m_{\eta 2}^2} \int \frac{d^3 q'}{(2 \pi)^3 2 q'^0} \int_{|q' - M_N^2/(4 q')|}^\infty \frac{dp}{2 \pi} \frac{p}{2 p^0} \int_{|q' - M_N^2/(4 q')|}^\infty \frac{d q}{2 \pi} \frac{q}{2 q^0} \\
	& \times \frac{1}{2 q'} \frac{1}{2 q'} \frac{(M_N^2 - 4 q'^2)(M_N^2 - 2 q' (p^0 + q^0))}{4 q'^2} \delta f_N (p) f_{\ell j} (q-q') f_N (q) \\
	=& 4 \text{Im} [Y_i^{(1)} Y_j^{(1)*} Y_j^{(2)} Y_i^{(2)*}] \frac{M_N^2}{m_{\eta 2}^2}  \frac{T^2}{32 \pi^3} \\
	& \times \int \frac{d q' q'}{2 \pi} \frac{(M_N^2 - 4 q'^2)(2 M_N^2 - 8 q' \sqrt{(q' - M_N^2/(4 q'))^2+M_N^2} - 6 q' T)}{64 q'^4} e^{- 3 \sqrt{(q' - M_N^2/(4 q'))^2+M_N^2}/T} e^{q'/T}.
\end{align*}
Since we assume $M_N \gg T$, we can approximate
\begin{equation}
	\sqrt{(q' - M_N^2/(4 q'))^2+M_N^2} \approx \frac{M_N^2}{4 q} + q.
\end{equation}
With this we can evaluate the above integral and obtain
\begin{equation}
	S_{\ell i}^\text{v} = 4 \text{Im} [Y_i^{(1)} Y_j^{(1)*} Y_j^{(2)} Y_i^{(2)*}] \frac{M_N^2}{m_{\eta 2}^2} \frac{T^2}{64 \pi^4} \frac{1}{32} (6 \sqrt{6} M_N T K_1 (\sqrt{6} M_N/T) + 4 M_N^2 K_0 (\sqrt{6} M_N/T)).
\end{equation}
Dividing by
\begin{equation}
	n_N = \frac{1}{2 \pi^2} M_N^2 T K_2 (M_N/T),
\end{equation}
and shifting to comoving coordinates, we find
\begin{equation}
	S_{\ell i}^\text{v} = \text{Im} [Y_i^{(1)} Y_j^{(1)*} Y_j^{(2)} Y_i^{(2)*}] \frac{M_N^2}{m_2^2} \frac{a_R}{M_N} \frac{1}{2^8 \pi^2} \frac{(6 \sqrt{6} K_1 (\sqrt{6} z)/z + 4 K_0 (\sqrt{6} z))}{K_2 (z)}.
\end{equation}

\clearpage

\bibliography{references}

\end{document}